# Kinematic classification of non-interacting spiral galaxies

Theresa Wiegert[a,b,*], Jayanne English[a]

[a]*University of Manitoba, Department of Physics & Astronomy, Winnipeg, MB, Canada, R3T 2N2*

[b]*Queen's University, Department of Physics, Engineering physics & Astronomy, Kingston, ON, Canada, K7L 3N6*

**Abstract**

Using neutral hydrogen (HI) rotation curves of 79 galaxies, culled from the literature, as well as measured from HI data, we present a method for classifying disk galaxies by their kinematics. In order to investigate fundamental kinematic properties we concentrate on non-interacting spiral galaxies. We employ a simple parameterized form for the rotation curve in order to derive the three parameters: the maximum rotational velocity, the turnover radius and a measure of the slope of the rotation curve beyond the turnover radius. Our approach uses the statistical Hierarchical Clustering method to guide our division of the resultant 3D distribution of galaxies into five classes. Comparing the kinematic classes in this preliminary classification scheme to a number of galaxy properties we find that our class containing galaxies with the largest rotational velocities has a mean morphological type of Sb/Sbc while the other classes tend to later types. Other trends also generally agree with those described by previous researchers. In particular we confirm correlations between increasing maximum rotational velocity and the following observed properties: increasing brightness in B-band, increasing size of the optical disk ($D_{25}$) and increasing star formation rate (as derived using radio continuum data). Our analysis also suggests that lower velocities are associated with a higher ratio of the HI mass over the dynamical mass. Additionally, three galaxies exhibit a drop in rotational velocity amplitude of $\gtrsim 20\%$ after the turnover radius. However recent investigations suggest that they have interacted with minor companions which is a common cause for declining rotation curves. (Figures 12, 14, 16 and 17 are interactive in the electronic pdf version of this paper.)

## 1. Introduction

The shape of a galaxy's rotation curve reveals more characteristics of a galaxy than a first glance will disclose. For example, the dark matter content and distribution in a galaxy is indicated by the flat rotation curve behaviour at outer radii [e.g. 5].

Using optical rotation curves, Persic et al. [58] developed a universal form for rotation curves (URC), dependent solely on the galaxy luminosity for 1100 galaxes of diverse Hubble types. The URC has however been debated by e.g. Bosma [6] since many galaxies do not fit the URC form [79], and most investigations point towards multi-parameter dependencies [e.g. 55] as a more likely scenario.

[*]Telephone: +1(613) 533-2719 Fax: +1(613) 533-6463

*Email addresses:* `twiegert@astro.queensu.ca` (Theresa Wiegert), `jayanne_english@umanitoba.ca` (Jayanne English)

Fig. 1 loosely corresponds to the rotation curve classes that were defined by Corradi and Capaccioli [21] in their kinematic classification scheme within optical radii (R25). We refer to these in section 4.1.

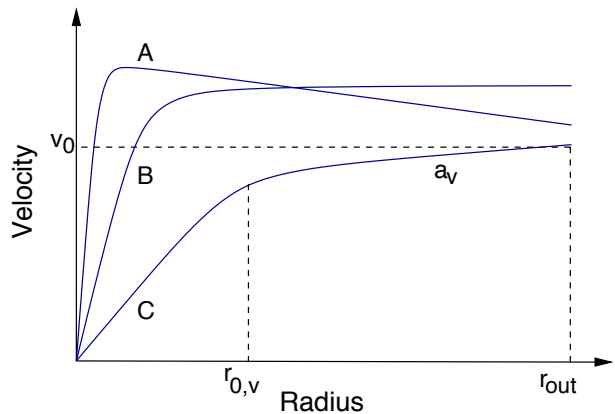

Figure 1: A schematic figure of (A) a rotation curve exhibiting steep rise/short turnover radius followed by a decrease, (B) a flat rotation curve and (C) a shallow rise/long turnover radius followed by a rise. Curve A also commonly exhibits a flat outer shape. We have overlaid the parameters described in § 3.1. The $r_{0,v}$ is the turnover radius, $v_0$ is the maximum velocity and $a_v$ is the asymptotic slope in Eq. 1.

Rotation curves from HI observations reveal the kinematic behaviour at larger radii, and still exhibit a flat shape. However there are exceptions to the schematic in Fig. 1 such as the rotation curve of NGC 7793, which has a turnover radius far from the galactic centre combined with a decreasing outer slope [14]. There are a few other cases where a subtle decrease can be perceived. The cause for this behaviour is still under debate: Are declining rotation curves a general trend for high surface brightness galaxies with short disk scale length [16] and galaxies with a big bulge? Or do they only appear when the galaxy is, or previously has been, interacting with another galaxy, which could result in tidal stripping of the halo (as has been shown to be the case for galaxies in clusters by Whitmore et al. [80])? Furthermore, Elmegreen and Elmegreen [28] report a correlation between declining curves and grand design spiral arm structures of galaxies.

Generally, there is a correlation between rotation curve shape and the central light concentration [e.g. 21, 17]. Galaxies with highly concentrated stellar light, such as S0 and Sab galaxies, have rotation curves with a rapid rise like curve A in Fig.1, reaching their turnover radius sooner (a few hundred parsecs from the centre) than galaxies with small bulges and relatively diffuse light distribution. In other words: dynamics in the central regions seem to be dominated by the stellar mass, as is also confirmed in HI studies [55]. This behaviour - a steep rise and a short turnover radius - is usually followed by a flat stretch at outer radii and is generally seen for high-luminosity galaxies. In contrast, low-luminosity galaxies exhibit rotation curves with a longer turnover radius and a rising curve at outer radii [65], implying that the dark matter fraction in galaxies increases as luminosity decreases. A number of authors have explored whether kinematic behaviour correlates with morphological Hubble type, but the answers are inconsistent: [e.g. 65, 10, 24] found no correlation, while others [e.g. 21] did find a correlation, although it is suggested only the inner region (e.g. bulge) have an influence [e.g. 55].

Chattopadhyay and Chattopadhyay [17] used principal component analysis to deduce which characteristics are more important in affecting the rotation curve shape. They studied the impact of, for example, absolute blue magnitude, disk scale length, central density of the halo, and presence of bar. Their most notable results show that while the overall shape is determined by the central surface brightness, the outer rotation curve (beyond the stellar disk) is mainly determined by the size of the stellar disk.

To summarize, there are still a few ambiguities and/or uncertainties associated with putative correlations of galaxy features with rotation curve shape. Apart from the luminosity and central light concentrations are there trends with other features? How pronounced is the association with rotation behaviour and morphology? What situations could cause a declining rotation curve? A number of reasons can contribute to this. For example, many of the investigations using HI data have small numbers of galaxies, leading to low number statistics. Additionally, often a particular galaxy group is targeted, such as late-type and low-luminosity or dwarf galaxies [72, 13]. Noordermeer et al. [55] provide one of few investigations that target early-type galaxies. To our knowledge, there are no HI investigations that make comparisons using a sample of isolated, non-interacting galaxies alone. Such systems provide an opportunity to study intrinsic kinematic and morphological correlations without the complicating effects of galaxy-galaxy interactions.

A kinematic classification scheme based on HI rotation curves from non-disturbed galaxies, could help illuminate commonalities between galaxies of similar kinematics. Such a classification scheme could be used to search for relationships between the observables of the galaxies. Also, provided with only the shape of the rotation curve of a galaxy, a kinematic classification scheme would provide information on which characteristics are most commonly associated with this shape. This would for example be useful for picking galaxies for projects aiming to investigate galaxies with specific characteristics. Once kinematic classes are tested observationally and reasonably established, template rotation curves could be developed for matching to observational data. For example one could combine together 3-D cubes of galaxies in order to create a high signal-to-noise "atlas" of features. This would be analogous to "stacking" spectra together and could similarly assist in classification, in this case, of faint or high redshift galaxies and in estimating their characteristics. Since galaxies with similar kinematics need to be combined into a 3-D template cube, a kinematic classification scheme would play an important role in the selection of input galaxies.

In this paper, we investigate whether it is possible to define a galaxy classification scheme, based on a number of kinematic parameters associated with the rotation curve. We present a potential candidate scheme based on our findings. Our method employs a mathematical approach, [hierarchical clustering, 37] combined with visual evaluation, to find features in a database of rotation curve properties. For the division into rotation curve groups (classes), we take into account parameters describing the observed velocity behaviour, such as the maximum velocity, the turnover radius, and the behaviour of the curve at large radii assessed by the rotation curve slope. These divisions are then examined for possible correlations with "secondary" parameters, such as morphology, disk size, star formation rate, and other galaxy characteristics.

In order to discern the intrinsic properties of rotation curves, emphasis is placed on using data from isolated non-interacting galaxies. Note that we focus on the outer rotation curve shape in this paper, adopting only one parameter (turnover radius) to define the inner region.

§ 2 describes the acquisition of rotation curves used for the classification and § 3.1 presents the parametrised rotation curve. § 3.2 describes the classification procedure using a clustering algorithm. The kinematic classes and uncertainties are described in § 4.1 and § 4.2. The search for correlations between the classification scheme and a number of additional galaxy characteristics is described in § 5 and the results are summarized in § 5.8. The discussion in § 6 compares the results with proposed trends from previous literature and § 7 provides our conclusions.

## 2. Data

In this section, we discuss the observational datasets from which we extracted rotation curves (§ 2.2) and describe our selection of rotation curves from those presented in the literature (§ 2.1). The total of 79 spiral galaxies (Sa through Sd) are sufficiently isolated that they are apparently without a history of interaction, which might have substantially altered their intrinsic rotational behaviour. The parametric rotation curve form, fitted to all the rotation curve data in order to extract kinematic parameters, is described in Section 3.1.

*2.1. Rotation curves from the literature*

In addition to 14 rotation curves generated in Wiegert [81] (§ 2.2), 74 HI rotation curves (some for the same targets) were culled from the literature to increase the total sample size. In order to have a sufficient resolution to measure rotation curve parameters, the maximum distance to these galaxies was 80 Mpc. The adoption of specific distances is described in § 5. Galaxies described by authors as interacting or distorted were not included. The resultant set consists of rotation curves from 79 galaxies. Some galaxies appear both in literature and in the data used in Wiegert [81], making it possible to assess uncertainties in the extracted rotation curve parameters.

| Reference | # of RCs | Comments |
|---|---|---|
| Deul and van der Hulst [26]* | 1 | M33 |
| Puche et al. [61]* | 1 | NGC 55 |
| Carignan and Puche [15]8 | 1 | NGC 247 |
| Puche et al. [60]* | 1 | NGC 300 |
| Casertano and van Gorkom [16]* | 1 | NGC 2683 |
| Meurer et al. [52]* | 1 | NGC 2915 |
| Broeils [8]* | 5 | |
| Begeman [3]* | 2 | |
| Cote et al. [22]* | 1 | NGC 5585 |
| Sancisi and van Albada [66]* | 1 | NGC 5907 |
| van der Hulst et al. [75]* | 1 | UGC 128 |
| Carignan et al. [12]* | 1 | NGC 6946 |
| Roelfsema and Allen [63]* | 1 | UGC 2885 |
| García-Ruiz et al. [31] | 18 | Edge-on galaxies |
| Sanders and Verheijen [68] | 23 | Ursa Major galaxies |
| de Blok et al. [23] | 13 | The THINGS survey |
| Hoekstra et al. [34] | 3 | |
| Bottema and Verheijen [7] | 1 | NGC 3992 |
| Fraternali et al. [29] | 1 | NGC 891 |
| Wiegert [81] | 14 | GalAPAGOS rotation curves |

Table 1: Rotation curves were extracted from these samples. Note that a galaxy may appear more than once in different samples, so the total number of individual galaxies (79) is less than the total in this list. Rotation curves from references with an asterisk were culled from the rotation curve collection in Sanders [67].

The literature describes 42 of the galaxies in our analysis as isolated or non-interacting. We designate these as Sample 1 (see Table 3). The 17 galaxies of Sample 2 (see Table 4) are well documented, yet lack mention of their isolation or interaction status, and were deemed to be potentially non-interacting even if a nearby companion or a slightly asymmetric morphology was mentioned in a few cases. We note that if there are asymmetries in a galaxy's rotation curve, this might be an echo of the galaxy's recent interaction history making it possible to distinguish an isolated galaxy from an interacting one. However, a large number (up to 50%) of apparently isolated field galaxies can show significant rotation curve asymmetries and lopsided HI profiles [e.g. 33, 62]. Thus it is unclear that asymmetries in the rotation curve are due to (recent) interaction, and the galaxies in Sample 2 exhibiting asymmetries may still be valid for our exploration. The remaining galaxies lacked sufficient information about interactions. Based upon their seemingly un-disturbed morphology and lack of companions, these 20 were collected into Sample 3 (see Table 5).

*2.2. Modelled galaxies*

For 14 galaxies, listed in Table 2, we determined our own rotation curves from neutral hydrogen spectral line datasets that were in the public domain (THINGS VLA data [23]) or donated (J. Irwin VLA data [45]; R. Swaters WHISP data [76]). These datacubes were modelled using the parametric galaxy modelling software currently called GalAPAGOS (under development by J. Fiege and the authors). An additional two galaxies were modelled with GalAPAGOS: NGC 55 [HIPASS, 46] and NGC 3621 (THINGS). Their results were used in uncertainty calculations (see § 4.2). However the rotation curves from literature were used for the other investigations in this paper since the NGC 55 data are of low resolution and the irregular emission in the outer disk of NGC 3621 makes our modelling unreliable.

While an updated version of GalAPAGOS will be described in Fiege et al. (in progress), the rotation curves used in this paper were analysed in the PhD thesis by Wiegert (2011)[1] and the 2010 GalAPAGOS version is described therein. Here we use the results from the rotation curve models that are labelled "best" in Table 6.10 of Wiegert's thesis. That is, GalAPAGOS generates a family of solutions within $1\sigma$ of the noise of each HI cube. A comparison is made between these resultant models and the data using a standard reduced $\chi^2$ analysis. The model with the lowest reduced $\chi^2$ was designated as "best".

Additionally the availability of Sloan Digitized Sky Survey i-band data [82] for a subset of 10 of these galaxies, marked with an asterisk in Table 2, allowed the generation of photometric luminosity profiles and mass models, detailed in Wiegert's thesis, which will be presented in other papers. Here we use the i-band scale lengths determined using the GIPSY task ELLINT and fitting the profile with an exponential disk and a Sérsic bulge.

| **Galaxy** | **Data** | **D (Mpc)** | **ref** | **DSL (kpc)** |
|---|---|---|---|---|
| (1) | (2) | (3) | (4) | (5) |
| NGC 55 | H | 1.6 | (1) | |
| NGC 925 | T | 9.12 | (2) | |
| NGC 2403* | T,S | 3.18 | (3) | 1.6 |
| NGC 2841* | T,S | 14.1±1.5 | (10) | 4.4 |
| NGC 2903* | W,T,I,S | 8.9 | (4) | 2.6 |
| NGC 3198* | T,S | 13.8+/-0.51 | (2) | 3.7 |
| NGC 3351* | T,S | 9.33 | (2) | 2.1 |
| NGC 3521* | T,S | 8.5 | (5) | 4.5 |
| NGC 3556* | I,S | 11.6 | (6) | 3.4 |
| NGC 3621 | T | 6.6 | (2) | |
| NGC 4096 | T | 10.2 | (7) | |
| NGC 4258* | W,S | 7.83 | (2) | 8.7 |
| NGC 5055* | T,S | 7.2 | (8) | 4.0 |
| NGC 7331* | T,S | 14.52 | (2) | 6.4 |
| NGC 7793 | T | 3.9 | (9) | |

Table 2: 15 galaxies for which rotation curves were derived using GalAPAGOS in Wiegert's thesis. Column 1: Galaxy name, where asterisks denote that luminosity profiles and mass models were derived for these galaxies in Wiegert [81], Column 2: H=HI cube from the HIPASS survey [46], T=HI cubes from the THINGS survey [23], W=HI cubes from the WHISP survey [76], S=i-band data from the Sloan Digital Sky Survey (SDSS) [82], I= HI cube provided by J. Irwin. Column 4 lists the references for the distances in Column 3: (1) Puche et al. [61], (2) Freedman et al. [30], (3) Madore and Freedman [49], (4) Karachentsev et al. [41], (5) Zeilinger et al. [84], (6) King and Irwin [45], (7) García-Ruiz et al. [31], (8) Pierce [59], (9) Karachentsev [39], (10) Macri et al. [48]. Column 5 lists the disk scale lengths derived from the SDSS data in Wiegert (2011) for those galaxies available in SDSS.

---

[1]Available at https://mspace.lib.umanitoba.ca/handle/1993/4377

# 3. Analysis

## 3.1. A parametric form of the rotation curve

A galaxy's rotation curve can be parameterized in the following manner:

$$v = v_0 \tanh\left(\frac{r}{r_{0,v}}\right)\left[1 + a_v\left(\frac{r}{r_{out}}\right)\right] \quad (1)$$

Eq. (1) is the same rotation curve form that has been used in the HI modelling software GalAPAGOS (2010) and includes four parameters; the maximum rotational velocity $v_0$, the turnover radius $r_{0,v}$, the slope $a_v$ of the rotation curve beyond the turnover radius, and the outer radius, $r_{out}$, which we define as the last observed data point for the analysis of the rotation curves from the literature, see Fig. 1. For the rotation curves derived using the GalAPAGOS software, $r_{out}$ is the radius at which the extrapolation of the outer gradient attains zero emission. To make all the rotation curves individually comparable, we scale the turnover radius by the galaxy radius $r_{out}$. This leaves us with a parameter set of three dimensions to use in the hierarchical clustering analysis employed to construct a classification scheme.

The values for the rotational velocity as a function of radius were carefully extracted by hand from published plots for all the galaxies in the samples. The analytical form of the rotation curve in Eq. (1) was fitted to these data, using a simple unconstrained linear optimization method, *fminsearch*, available in MATLAB. This fitting procedure finds the best fit of a function of several variables to a number of data point values. The resulting parameters – $v_0$, $r_{0,v}$ and $a_v$ – are shown in Tables 3, 4, and 5 for samples 1, 2 and 3 respectively.

Fig. A.21 in Appendix shows 74 of the extracted rotation curves with the analytical rotation curve fits overlaid. Fig. 2 shows three examples of the rotation curve fits. 12 rotation curves from the literature are included in Fig. A.21 for comparison with those from GalAPAGOS in order to assess uncertainties in § 4.2. The remaining five rotation curves derived from HI data cubes (§ 2.2) do not have counterparts in the literature and so are not included in Fig. A.21.

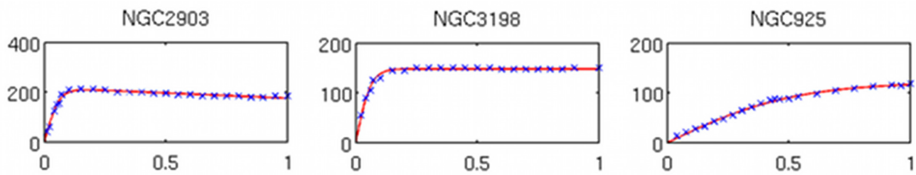

Figure 2: Three examples of the rotation curve fits in Appendix A, showing the three types of rotation curves in Fig. 1.

## 3.2. Hierarchical clustering on rotation curve parameters

A clustering algorithm is capable of taking many parameters into account simultaneously and search for commonalities in the data in order to divide them into groups, also called clusters. There are different techniques, for example, in how the mathematical 'distance' between parameters is defined, as well as in interpretation of the results.

Hierarchical clustering, which in iterative steps merges similar data points into clusters, is a suitable method for this investigation, due to its flexibility and ease of interpretation. Its display of the results in a hierarchical tree is especially useful for finding subcluster structures and discerning relationships.

The basic process of hierarchical clustering (based on Johnson [37]) on a set of N parameters (i.e. N-dimensions) is the following:

1. Each item in the set is assigned to its own cluster. Thus the set contains N clusters, each containing just one item. In this case, the distance in the parameter space between a pair of clusters, is equal to the distance between the pair of items.
2. The most similar (i.e. the closest) pair of clusters is found by calculating all the distances in parameter space. These two clusters are merged into a single cluster, diminishing the number of clusters by one.
3. The distances are recalculated between the new cluster and each of the old clusters. Note that the clusters at each iteration are monitored, enabling the algorithm to return a tree structure of the clusters present at each iteration.
4. The second and third steps are now repeated until all the items are clustered into a single cluster containing all N items.

An example of clustering five datapoints in two dimensions is shown in Fig. 3.

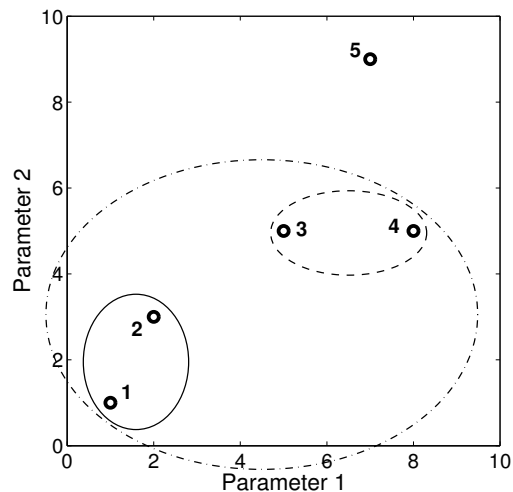

Figure 3: A schematic figure of how the function of hierarchical clustering. Data points 1 through 5 are clustered based on the distances between the points. The first clusters consist of each point individually. The second cluster consists of point 1 and 2 (solid ellipse), and the third cluster consists of points 3 and 4 (dashed ellipse). The fourth cluster consists of the 2nd and 3rd cluster (points 1-4, dash-dotted ellipse) and the last cluster consists of all 5 points in the set. The sizes of the ellipses are arbitrary in this figure and are only used to illustrate the clusters.

The graphical representation of the resulting hierarchy structure, a *dendrogram*, is a reversed tree structure where all items branch together to form one single cluster. Fig. 4 shows the dendrogram of the clusters in Fig. 3. This two dimensional example can be extended to any number of dimensions. Here, we only needed to extend our parameter space to three dimensions, as described in § 3.1.

We have used the MATLAB statistics toolbox which includes functions that can perform hierarchical clustering on an N parameter dataset.

After having normalised the extracted rotation curve parameters so all galaxies would be treated on an equal footing, the hierarchical clustering was performed by letting a script feed the parameters into the *pdist* function. Pdist calculated the Euclidian distances between the data points after each call of the *linkage* function that links items into clusters. The entire linkage tree was displayed using the function *dendrogram*. The results were then analysed, using tools in MATLAB's Statistics Toolbox, as well as visual inspection.

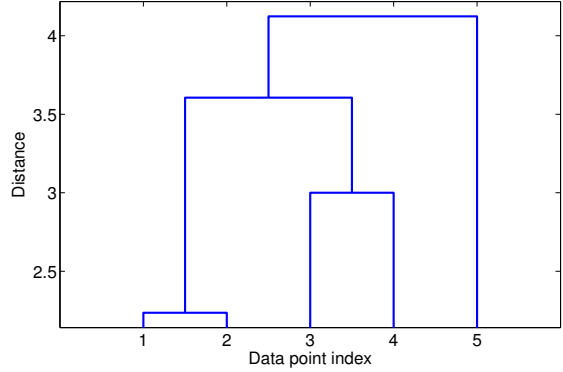

Figure 4: A dendrogram visualizing the cluster tree. The numbers along the horizontal axis are the indices of the data points shown in Fig. 3. The data are connected by lines shaped as an upside-down U. The height of the U-shape represents the Euclidian distance between the objects in parameter space, given by the values on the vertical axis.

The algorithm attempts to find all naturally occurring clusters in the data set. However, there are cases where a user may justify defining a more sparsely populated area as a group with some commonalities, which the automated process might have difficulties finding. A careful visual inspection is therefore an important step in the clustering process, in order to confirm any natural clusters as well as potential subclusters.

## 4. Results

As expected, given that the number of rotation curve shapes are limited (e.g. Fig. 1), the 3D volume defined by the three kinematic parameters is populated mainly in specific regions. This is apparent in Fig. 5, which shows 3 views of this ($v$, $r_{0,v}$, $a_v$) volume. Fig. 5 also indicates that the distribution of galaxies within this volume may form a continuous 3D shape. One option for a scheme is to devise a labelling that simply specifies the location of a galaxy within this shape and thereby use its position in the volume to indicate the galaxy's characteristics. However some analyses, such as a search for correlations, can benefit from segmenting a continuous distribution into classes. This approach is common and exemplified by the Hubble Tuning Fork diagram. The results of our attempt to delineate kinematic classes are described in § 4.1 while the robustness of the adopted classes is explored in § 4.2. We look for correlations between the kinematic classes and other galaxy characteristics in the next section, § 5.

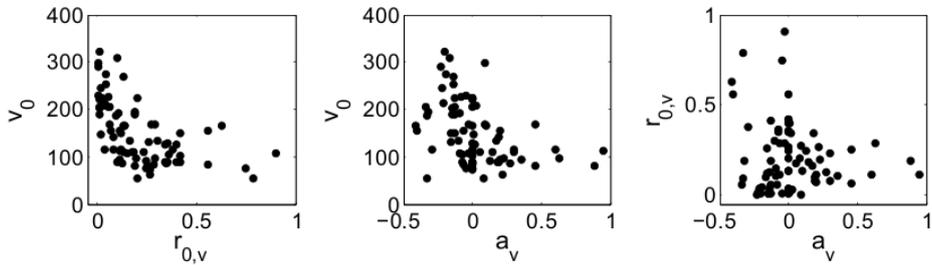

Figure 5: Three views of the total sample of galaxies distributed throughout the parameter space of velocity amplitude ($v_0$), slope ($a_v$), and turnover radius ($r_{0,v}$). The 3D distribution is shown in Fig. 7 with a rotatable version available in Fig. 12.

*4.1. Kinematic classification scheme*

Rather than relying on visual inspection alone, we have used the hierarchical clustering method in § 3.2 to explore whether three rotation curve parameters can delineate a kinematic classification scheme. The extracted values of the parameters (turnover radius, maximum velocity, and slope) are listed in Tables 3, 4 and 5 for the three samples, described in Section 2, along with references. The hierarchical clustering script was applied to the galaxies from all three samples together as well as to the more clearly non-interacting galaxies in Sample 1 alone. Below we describe the output that guided us in the adoption of the five clusters which we specify as kinematic class designations in Tables 3, 4 and 5.

Fig. 6 (the dendrogram) and Fig. 7 display the result of the clustering algorithm applied to the full set of 79 galaxies. Although no natural, e.g. well separated, clusters can easily be discerned (and may not exist if the distribution within the $v_0 - a_v - r_{0,v}$ volume is continuous), we adopt a division into groups based on the result. When displaying $v_0$, $a_v$ and $r_{0,v}$ (Fig. 7), we can also visually delineate the clusters we have adopted, in three dimensions. The structure leading to our scheme becomes particularly apparent when rotating the 3D Fig. 12 in the pdf version of this paper.

In both the dendrogram and 3D displays the most distinct cluster consists of high rotational velocity galaxies. We have designated this group A, where 'A' stands for the amplitude of the velocity. Slightly lower velocities tend to spread out towards higher values of turnover radius and slope, separated by the algorithm into two groups which we designate as AS and AT ('S' for slope and 'T' for turnover). There are a few low velocity galaxies and these visually suggest the presence of two 'branches' in the 3D plot, one for rotation curves with high slopes and one with high turnover radius values. These branches are not output as a "natural" cluster by the clustering functions, since the points are widely spread across these regions of parameter space. Nevertheless, due to the ability to visually distinguish the branches in the 3D view Fig. 12, we designate them T for the turnover radius branch and S for the slope branch.

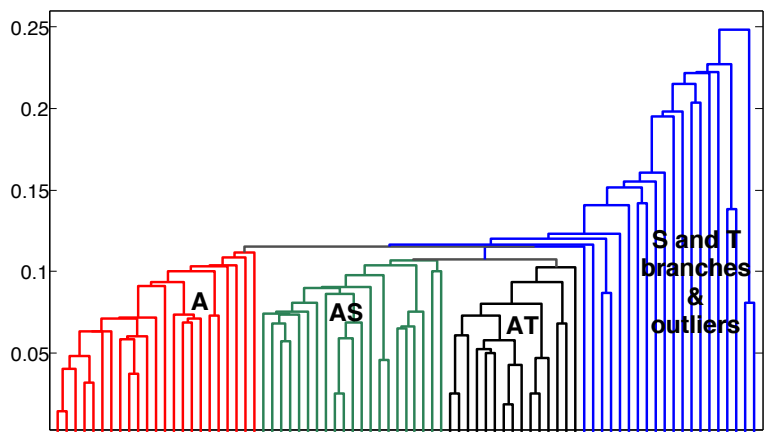

Figure 6: A cluster tree (or dendrogram) showing the groups of similarities between the three rotation curve parameters of the 79 galaxies in the sample and the division we have adopted, guided by the dendrogram. For example, Class A (red connectors) consists of galaxies with high rotational velocities, low slope values and short turnover radii. See text for the other designations. Note that two branches on the very right shown among the outliers (blue connectors) are more apparent in Fig. 7.

Fig. 8 and Fig. 9 show the 3D figure and the cluster tree of the 42 confirmed non-interacting galaxies in Sample 1. With the smaller sample, only one of the outliers exists. However the overall shape of the distribution is the same and each cluster is represented in this less populated version. Although we examine the robustness further in § 4.2, this figure encourages us to adopt the kinematic class scheme devised using all galaxies from the three samples, and the rest of the analysis is performed on the full set of galaxies.

| Name | ref | $v_0$ | $a_v$ | $r_{0,v}$ | $r_{out}$ | Kin. class |
|---|---|---|---|---|---|---|
| M33 | (a) | 115.16 | 0.6030 | 0.1158 | 9.0 | S |
| NGC 55 | (b) | 86.35 | -0.0011 | 0.3484 | 9.8 | AT |
| NGC 247 | (c) | 112.76 | 0.9471 | 0.1153 | 11 | S |
| NGC 300 | (d) | 95.87 | 0.2280 | 0.1949 | 12 | AS |
| NGC 891 | (e) | 228.86 | -0.0431 | 0.0061 | 19 | A |
| NGC 925 | (f,g) | 126 | 0.0200 | 0.3400 | 2.2 | AT |
| NGC 2403 | (f,g) | 141 | 0.1900 | 0.0800 | 22 | AS |
| NGC 2613 | (g) | 307.5 | -0.1700 | 0.1000 | 43 | A |
| NGC 2683 | (h) | 205.41 | -0.3389 | 0.0575 | 18 | A |
| NGC 2841 | (f,g) | 322 | -0.2000 | 0.0100 | 60 | A |
| NGC 2903 | (f,g) | 206 | -0.1300 | 0.0600 | 27 | A |
| NGC 2915 | (i) | 88.98 | 0.1893 | 0.0981 | 15 | AS |
| NGC 2976 | (f) | 82.62 | 0.0016 | 0.5557 | 2.5 | T |
| NGC 2998 | (j) | 208.41 | 0.0267 | 0.0325 | 47 | A |
| NGC 3198 | (f,g) | 155 | -0.0700 | 0.1100 | 37 | AS |
| NGC 3351 | (f,g) | 202.2 | -0.1500 | 0.0100 | 17 | A |
| NGC 3510 | (k) | 88.17 | 0.0020 | 0.4001 | 4.7* | AT |
| NGC 3521 | (f,g) | 245 | -0.2200 | 0.0140 | 38 | A |
| NGC 3556 | (g) | 154.6 | -0.4000 | 0.5600 | 20 | T |
| NGC 3621 | (f,g) | 155.35 | 0.1958 | 0.0704 | 26 | AS |
| NGC 3877 | (l,m) | 168.92 | -0.0302 | 0.2881 | 9.8 | O |
| NGC 3917 | (l,m) | 134.12 | -0.0983 | 0.3035 | 13 | AT |
| NGC 3972 | (l,m) | 128.78 | -0.0016 | 0.3563 | 7.5 | AT |
| NGC 3992 | (l,m) | 267.89 | -0.1417 | 0.1304 | 30 | A |
| NGC 4096 | (g) | 147 | 0.0120 | 0.0160 | 9.2 | AS |
| NGC 4100 | (l,m) | 194.58 | -0.3262 | 0.1908 | 20 | A |
| NGC 4138 | (l,m) | 186.38 | -0.3271 | 0.0955 | 16 | A |
| NGC 4144 | (k) | 73.59 | -0.0009 | 0.2671 | 4.9* | AT |
| NGC 4258 | (g) | 206 | 0.0000 | 0.0110 | 25 | A |
| NGC 5033 | (n) | 224.48 | -0.1270 | 0.0097 | 34 | A |
| NGC 5055 | (f,g) | 212 | -0.2100 | 0.0345 | 32 | A |
| NGC 5371 | (n) | 221.53 | 0.0017 | 0.0124 | 40 | A |
| NGC 5533 | (j) | 289.22 | -0.2282 | 0.0041 | 72 | A |
| NGC 5585 | (o) | 92.3 | 0.1272 | 0.2442 | 12 | AS |
| NGC 5907 | (p) | 226.05 | -0.0807 | 0.0577 | 32 | A |
| NGC 6503 | (q) | 115.77 | 0.0030 | 0.0390 | | AS |
| NGC 6674 | (j) | 272.88 | -0.1891 | 0.0443 | 70 | A |
| NGC 7331 | (f,g) | 254 | -0.1400 | 0.0400 | 3.5 | A |
| NGC 7793 | (f,g) | 115 | -0.29 | 0.3800 | 12 | AT |
| UGC 128 | (r) | 132.79 | 0.1702 | 0.1638 | 40 | AS |
| UGC 7089 | (k) | 79.92 | 0.8858 | 0.1889 | 6.2* | S |
| UGC 9242 | (k) | 96.61 | 0.6257 | 0.2892 | 10* | S |

Table 3: Sample 1 – 42 galaxies known to be isolated and/or non-interacting. The references in column 2 refer to the paper with the originally published rotation curves which were used to derive the three rotation curve parameters in columns 3 (velocity amplitude), 4 (rotation curve slope) and 5 (turnover radius). References in column 2 (Tables 3, 4 and 5): (a) Deul and van der Hulst [26], (b) Puche et al. [61], (c) Carignan and Puche [15], (d) Puche et al. [60], (e) Fraternali et al. [29], (f) de Blok et al. [23], (g) Wiegert [81], (h) Casertano and van Gorkom [16], (i) Meurer et al. [52], (j) Broeils [8], (k) García-Ruiz et al. [31], (l) Sanders and Verheijen [69], (m) Verheijen [79], (n) Begeman [3], (o) Cote et al. [22], (p) Sancisi and van Albada [66], (q) Hoekstra et al. [34], (r) van der Hulst et al. [75], (s) Carignan et al. [12], (t) Roelfsema and Allen [63]. Column 6 lists the outer radius (kpc), to which the turnover radius was normalised. Outer radii listed with an asterisk are only available in angular size in the García-Ruiz et al. [31] sample and have here been converted to kpc using distances in column (3) of Table 9. Galaxies with no given outer radius are from the Hoekstra et al. [34] sample, where no outer radii were provided. See § 4.1 for a description of kinematic classes listed in column 7.

| Name | ref | $v_0$ | $a_v$ | $r_{0,v}$ | $r_{out}$ | Kin. class |
|---|---|---|---|---|---|---|
| NGC 0801 | (j) | 223.17 | 0.0019 | 0.0245 | 60 | A |
| NGC 1003 | (j) | 115.28 | 0.3008 | 0.0784 | 32 | AS |
| NGC 1560 | (q) | 80.29 | 0.4549 | 0.2505 | | S |
| NGC 2770 | (k) | 165.71 | -0.4147 | 0.6249 | 12* | T |
| NGC 3600 | (k) | 94.81 | 0.2039 | 0.1111 | 14* | AS |
| NGC 3726 | (l,m) | 166.58 | 0.1015 | 0.1328 | 28 | AS |
| NGC 3949 | (l,m) | 169.34 | 0.0862 | 0.2727 | 6.1 | O |
| NGC 3953 | (l,m) | 223.46 | -0.0007 | 0.2045 | 14 | O |
| NGC 4010 | (l,m) | 126.23 | 0.0047 | 0.3980 | 9.0 | AT |
| NGC 4013 | (l,m) | 190.19 | -0.1464 | 0.0092 | 27 | A |
| NGC 4051 | (l,m) | 168.41 | 0.4595 | 0.0626 | 11 | O |
| NGC 4217 | (l,m) | 189.76 | -0.1266 | 0.1872 | 14 | A |
| NGC 5301 | (k) | 149.75 | -0.1273 | 0.4155 | 14* | O |
| NGC 6946 | (s,f) | 164.84 | -0.0010 | 0.1372 | 30 | AS |
| UGC 2459 | (k) | 149.77 | -0.1547 | 0.1204 | 53* | AS |
| UGC 2885 | (t) | 299.16 | 0.0912 | 0.0026 | 72 | O |
| UGC 3909 | (k) | 76.31 | -0.0027 | 0.2458 | 14* | AT |

Table 4: Sample 2 – 17 well-documented galaxies with no mention of interaction. These are placed in Sample 2 due to an apparent companion or asymmetry, which might be signs of interaction. The references in column 2 (see caption of Table 3) refer to the paper with the originally published rotation curves which were used to derive the three rotation curve parameters in columns 3 (velocity amplitude), 4 (rotation curve slope) and 5 (turnover radius). Column 6 lists the outer radius (kpc), to which the turnover radius was normalised. Galaxies with no given outer radius are from the Hoekstra et al. [34] sample, where no outer radii were provided. See § 4.1 for a description of kinematic classes listed in column 7.

| Name | ref | $v_0$ | $a_v$ | $r_{0,v}$ | $r_{out}$ | Kin. class |
|---|---|---|---|---|---|---|
| NGC 3118 | (k) | 101.52 | -0.0026 | 0.4195 | 9.5* | AT |
| NGC 4157 | (m,l) | 192.49 | -0.0961 | 0.1078 | 26 | A |
| NGC 4183 | (m,l) | 109.91 | 0.0000 | 0.1336 | 18 | AS |
| NGC 4389 | (m,l) | 107.87 | -0.0287 | 0.9027 | 4.6 | T |
| NGC 5023 | (k) | 84.38 | -0.0638 | 0.2864 | 10* | AT |
| NGC 5229 | (k) | 53.97 | 0.0786 | 0.2037 | 9.6* | AT |
| UGC 1281 | (k) | 54.65 | -0.3282 | 0.7850 | 4.5* | T |
| UGC 2259 | (q) | 93.25 | 0.3538 | 0.1099 | | AS |
| UGC 3137 | (k) | 106.84 | -0.0885 | 0.1509 | 59* | AS |
| UGC 5459 | (k) | 130.93 | -0.1604 | 0.2584 | 18* | AT |
| UGC 6399 | (m,l) | 87.85 | 0.0010 | 0.4170 | 6.8 | AT |
| UGC 6446 | (m,l) | 85.11 | 0.2697 | 0.1251 | 13 | AS |
| UGC 6667 | (m,l) | 87.7 | 0.1776 | 0.3415 | 6.8 | O |
| UGC 6818 | (m,l) | 74.52 | -0.0441 | 0.7455 | 6.0 | T |
| UGC 6917 | (m,l) | 110.87 | 0.2965 | 0.2333 | 9.0 | AS |
| UGC 6930 | (m,l) | 110.38 | 0.1485 | 0.1817 | 14 | AS |
| UGC 6983 | (m,l) | 108.96 | 0.0492 | 0.1737 | 14 | AS |
| UGC 7321 | (k) | 104.07 | -0.0781 | 0.3621 | 12* | AT |
| UGC 7774 | (k) | 88.47 | -0.0721 | 0.3510 | 14* | AT |
| UGC 8246 | (k) | 62.78 | 0.2194 | 0.2648 | 10* | O |

Table 5: Sample 3 – these 20 galaxies lack sufficient information to determine conclusively whether they really are isolated and non-interacting. The references in column 2 (see caption of Table 3) refer to the paper with the originally published rotation curves which were used to derive the three rotation curve parameters in columns 3 (velocity amplitude), 4 (rotation curve slope) and 5 (turnover radius). Column 6 lists the outer radius (kpc), to which the turnover radius was normalised. Galaxies with no given outer radius are from the Hoekstra et al. [34] sample, where no outer radii were provided. See § 4.1 for a description of kinematic classes listed in column 7.

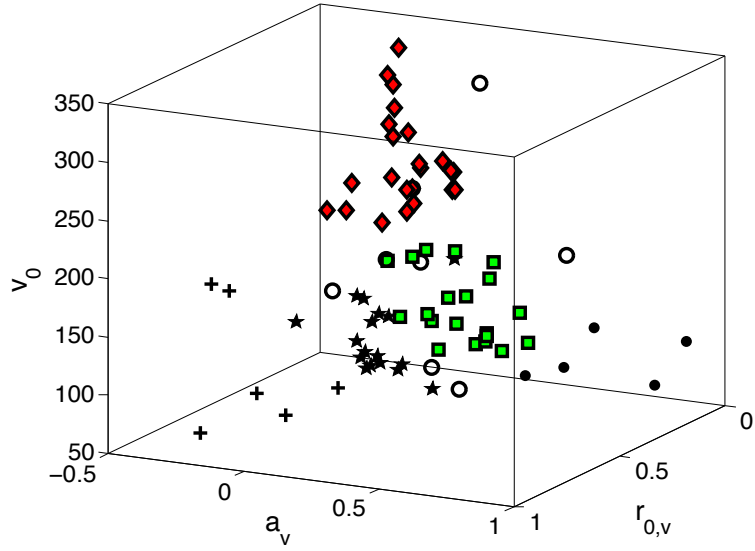

Figure 7: Division of galaxies into kinematic classes. Red diamonds = Class A, green squares = Class AS, black stars = Class AT, black dots = Class S, '+'-sign = Class T. Circles are outliers.

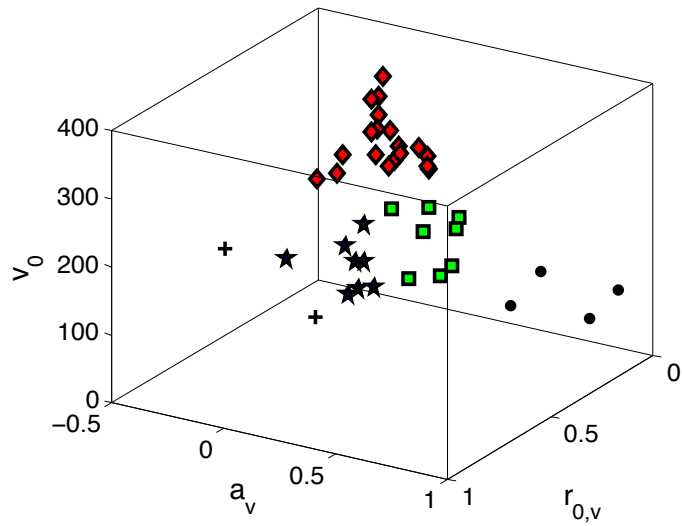

Figure 8: The three velocity parameters for each of the 42 galaxies documented to be isolated or non-interacting. The adopted classes are indicated by symbols: Red diamonds = Class A, green squares = Class AS, black stars = Class AT, black dots = S, '+'-sign = T.

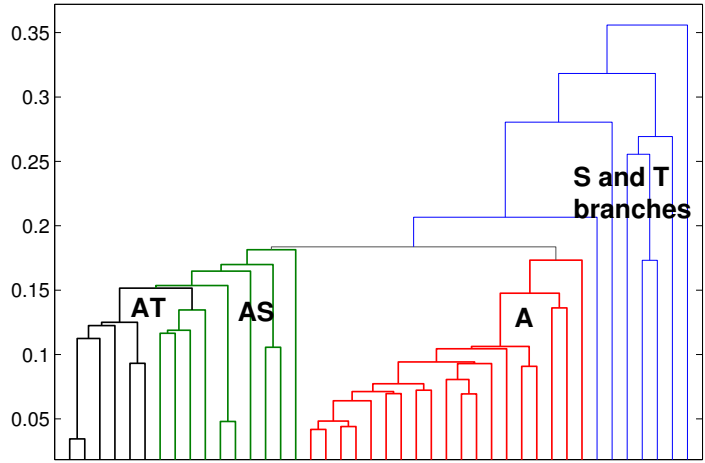

Figure 9: A cluster tree showing the similarities between the galaxies in Sample 1.The division of clusters are based on the results from the full sample.

The characteristic values for each class are listed in Table 6 and Fig. 10 shows the rotation curve fits of the galaxies plotted according to their different kinematic classes. Note that 15 of the rotation curves used for the analysis in this section were generated by GalAPAGOS. Specifically, ten GalAPAGOS rotation curves were used rather than their literature counterparts (Fig. A.21). We use the values in Table 6 along with the figures in this section to develop the synopsis of the kinematic classes presented here:

|  | $v_0$ (km/s) | | | $r_{0,v}(/r_{out})$ | | | $a_v$ | | |
| --- | --- | --- | --- | --- | --- | --- | --- | --- | --- |
| **Class** | min | max | mean | min | max | mean | min | max | mean |
| A | 186 | 322 | 229 | 0.004 | 0.19 | 0.05 | -0.34 | 0.03 | -0.15 |
| AS | 85 | 167 | 119 | 0.02 | 0.42 | 0.14 | -0.15 | 0.35 | 0.11 |
| AT | 54 | 134 | 100 | 0.20 | 0.42 | 0.33 | -0.29 | 0.08 | -0.04 |
| S | 80 | 115 | 97 | 0.12 | 0.29 | 0.19 | 0.45 | 0.95 | 0.70 |
| T | 55 | 166 | 107 | 0.56 | 0.9 | 0.7 | -0.41 | 0.002 | -0.20 |

Table 6: Minimum, maximum and mean values of the rotation curve parameters of the galaxies in the different kinematic classes.

**Class A** The most distinct group, consisting of fast rotating galaxies with short turnover radii and the lowest slope values. We note that 70% of the group have an $a_v$ value below -0.1, 30% below -0.2 and 13% below -0.3. None have strongly positive $a_v$ values.

To some readers it may appear in Fig. 10 that there are more rotation curves that are decreasing than the number (10) that are visually apparent in Appendix A for Class A galaxies. This is due to the inclusion of three galaxies that do not have counterparts in the literature and whose GalAPAGOS derived $a_v$ values are more negative than -0.14. Since there are 23 Class A galaxies, at least 57% of the curves in Fig. 10 contribute to the appearance that negative slopes dominate.

At this point we note that negative $a_v$ values are not necesserily synonymous with declining rotation curves, which we explore in § 5.7.

**Class AS** This group has lower velocity amplitudes than A, rotation curves that visually appear flat, and generally longer turnover radii than in A class.

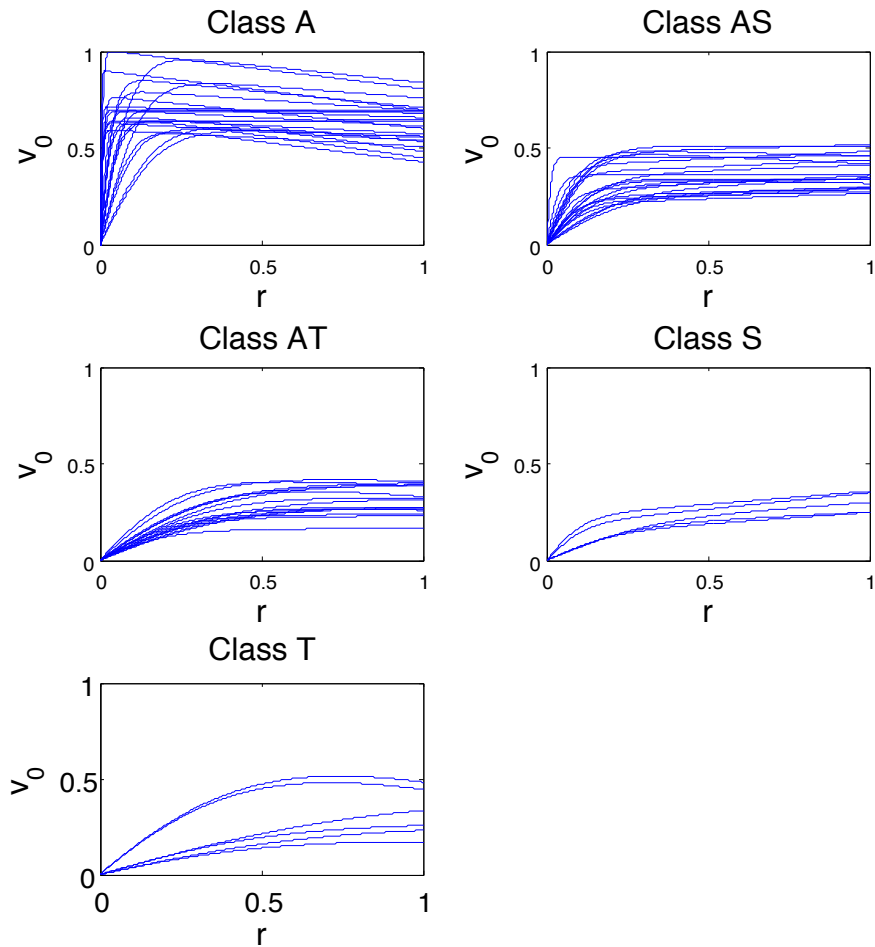

Figure 10: The rotation curve fits plotted according to their class designation. The radius ($r$) and the maximum rotational velocity ($v_0$) are normalised to values between 0 and 1.

**Class AT** The velocity amplitude range of this group spans a subset of class AS's velocity range. However the mean of the slope values is more negative and the mean of the turnover radii is longer. (In contrast to A, the turnover radii are the longest of the three classes.) Thus in the $a_v$- $r_{0,v}$ plane these galaxies tend towards a particular region of the 3D distribution. We present a figure displaying this region while discussing uncertainties in § 4.2.

**Class S** This "slope branch" could be interpreted as an extension to Class AS albeit with slightly lower velocity amplitudes. While it has intermediate turnover radii, it is distinguished by its highly increasing slope values.

**Class T** Similarly, this class may be an extension to Class AT since it has a similar velocity amplitude range. It has a couple of negative $a_v$ values, but the main distinguishing feature is the very large turnover radii of its galaxies.

**Outliers** The outliers are individually described in Table 7.

| Galaxy | $v_0$ (km/s) | $r_{(0,v)}$ | $a_v$ | Comment |
|---|---|---|---|---|
| NGC3877 | 169 | 0.29 | -0.03 | close to AS and NGC 3949. Classified as AT in the analysis using Sample 1 alone. |
| NGC3949 | 169 | 0.27 | 0.09 | close to AS and NGC 3877. |
| NGC3953 | 223 | 0.20 | 0 | A-like with high turnover radius value. |
| NGC4051 | 168 | 0.06 | 0.46 | High-velocity 'void' galaxy, close to S. |
| NGC5301 | 150 | 0.42 | -0.13 | AT-like, closer to T parameter space. |
| UGC2885 | 299 | 0.003 | 0.09 | A-like with an increasing slope. |
| UGC6667 | 88 | 0.34 | 0.18 | 'void', closer to AT parameter space. |
| UGC8246 | 63 | 0.26 | 0.22 | 'void', AS or AT, somewhat closer to S parameter space. |

Table 7: Eight galaxies classified as outliers. The "close" description refers to proximity to regions of the 3D parameter space associated with particular classes. The 'void' refers to the empty volume between the S branch and T branch, where galaxies with both a large turnover radius and an increasing slope would reside.

*4.2. Uncertainties*

Since the distribution of kinematic parameters is continuous, the boundaries between classes are not expected to be clearly delineated by galaxy properties alone. However the general regions of a classification scheme should not change drastically due to uncertainties in the measurement of the scheme's parameters. Therefore in this section we investigate the impact of such uncertainties.

One of the dominant uncertainties in measuring kinematic parameters is generated by different approaches to fitting rotation curves to data. In this work, although we fit the same equation, it is applied both to (1) plots of data points from the literature as well as to (2) all the data within 3D HI cubes. The literature rotation curves were initially generated with different modelling software such as ROTCUR in GIPSY. While GalAPAGOS weights voxels equally, the 1D fits to the literature curves may provide more detail on the inner regions due to applying equal weight at all radii. Also differences can be introduced due to, for example, the number of datapoints in the rotation curve and differing cut-offs for the rotation curves (due to different values of the HI radius $r_{out}$). Table 8 lists the galaxies in this study which allow us to compare the kinematic parameters derived using GalAPAGOS on HI cubes, in Wiegert [81], with those values derived from published rotation curves of the same galaxies. NGC 4096 values are derived from Westerbork Synthesis Radio Telescope datasets while NGC 55 additionally provides a comparison between HIPASS [46] and the VLA data [61] which were used for the clustering analysis. THINGS data [23] were used in both approaches (1) and (2) for the remainder of these galaxies.

**Outer radius** The radius of the HI disk is uncertain, in that it is hard to determine how individual authors have chosen this radius. Ideally, we would like to define $r_{out}$ to a certain column density similar to how the size of a galaxy is defined using a faint isophote for optical data (e.g. $D_{25}$). That is, we would select the furthest radius associated with a low, but specific, column density. Obviously this is not possible when relevant information is not available in the literature data.

|  |  | $v_0$ (km/s) |  | $a_v$ |  | $r_{0,v}$ (scaled) |  | $r_{out}$ (″) |  |  |
| --- | --- | --- | --- | --- | --- | --- | --- | --- | --- | --- |
| **Galaxy** | Ref | W | Lit | W | Lit | W | Lit | W | Lit | % |
| (1) | (2) | (3) | (4) | (5) | (6) | (7) | (8) | (9) | (10) | (11) |
| NGC 55   | a | 86  | 86  | 0.11  | 0.00  | 0.21 | 0.35 | 1330 | 1263 | 5 |
| NGC 925  | b | 126 | 116 | 0.02  | 0.00  | 0.34 | 0.52 | 500  | 291  | 42 |
| NGC 2403 | b | 141 | 137 | 0.19  | 0.19  | 0.08 | 0.08 | 1400 | 1160 | 17 |
| NGC 2841 | b | 322 | 316 | -0.20 | -0.16 | 0.01 | 0.04 | 872  | 512  | 41 |
| NGC 2903 | b | 206 | 210 | -0.13 | -0.20 | 0.06 | 0.06 | 658  | 672  | 2 |
| NGC 3198 | b | 155 | 148 | -0.07 | 0.00  | 0.11 | 0.06 | 547  | 568  | 4 |
| NGC 3521 | b | 245 | 225 | -0.22 | -0.28 | 0.01 | 0.09 | 915  | 598  | 35 |
| NGC 3621 | b | 146 | 155 | -0.09 | 0.20  | 0.08 | 0.07 | 728  | 812  | 12 |
| NGC 4096 | c | 147 | 149 | 0.01  | 0.00  | 0.05 | 0.22 | 186  | 210  | 13 |
| NGC 5055 | b | 212 | 208 | -0.20 | -0.20 | 0.03 | 0.05 | 928  | 980  | 6 |
| NGC 7331 | b | 254 | 253 | -0.14 | -0.12 | 0.04 | 0.06 | 493  | 351  | 29 |
| NGC 7793 | b | 115 | 113 | -0.29 | -0.34 | 0.38 | 0.27 | 440  | 410  | 7 |

Table 8: W refers to Wiegert's thesis and "Lit" refers to the literature referenced in column (2): a) Puche et al. [61] (VLA), b) de Blok et al. [23] (THINGS), c) Sanders and Verheijen [69], Verheijen [79] (WRST). Wiegert's analysis used THINGS data except NGC 4096 (WRST) and NGC 55 (HIPASS); see § 4.2. $v_0$ is half of the velocity amplitude, i.e. the maximum rotational velocity, $a_v$ is the slope of the rotation curve and $r_{0,v}$ is the turnover radius which is scaled by $r_{out}$, the outer radius. The last column (11) shows the difference in percentage between our measured value for $r_{out}$ and the value found in the literature.

The maximum deviation in the value of $r_{out}$ (up to 42% difference) occurs using THINGS data for both approaches for determining the rotation curves. While this may appear surprising, it arises from different definitions of $r_{out}$. The THINGS team uses an rms cutoff at the $3\sigma$ level (3 times the rms noise), which consequently discards faint emission in the galaxy outskirts for a few galaxies. In Wiegert's analysis, $r_{out}$ is one of many parameters fitted and delivered by GalAPAGOS (the 2010 version), the software we use to model the HI content of a galaxy, including emission in the outer regions. The fitted $r_{out}$ can thus be somewhat different from an observationally defined radius.

The rotation amplitude $v_0$ can be mildly affected by $r_{out}$. The asymptotic slope $a_v$ may have a large uncertainty if the difference between $r_{0,v}$ and $r_{out}$ is not large. However the largest impact is on the turnover radius $r_{0,v}$, since we have chosen to scale it by the outer radius. Thus an increase in $r_{out}$ leads to a decrease in $r_{0,v}$ of the same percentage.

**Turnover radius** The mean difference between our GalAPAGOS estimate and that from literature is 0.07 while the mode is 0.01. The larger difference (0.18) is associated with NGC 925 and is due to the effect on the turnover radius ($r_{0,v}$) by $r_{out}$. (The actual rotation curves from the 2 approaches are however very similar - see P. 117 in Wiegert's thesis.)

Obviously $r_{0,v}$ is also sensitive to the number of data points tracing the turnover in the rotation curve. NGC 55 represents the case where one of the datasets (HIPASS) has too few points. Although there are exceptions such as NGC 4013, for most galaxies in our study this region is well enough sampled to avoid this problem.

Another factor that could potentially have a large impact is the ability to find the centre of the galaxy while measuring the rotation curve from HI data, due to, for example, a paucity of HI in the centre. The difference between the rotation curves for NGC 2903 measured by Hoekstra et al. [34] and de Blok et al. [23] might suffer from this scenario, which could be partly responsible for the difference of almost 50% between their measured turnover radius values. While NGC 2841 and NGC 3521 in our comparison sample also have low HI in their central regions, the difference in their turnover values is not extremely different from the mean or mode. NGC 4096 represents an extreme case where neither rotation curve is reliable. The low signal-to-noise of the **Velocity** We find that the value of maximum velocity $v_0$ generally varies by about 3%. However ocassionally the difference in $v_0$ cited in the literature could be high. As an example of how we selected values if this was the case, we use NGC 6946 whose $v_0$ differs by 19% for the studies by Carignan et al. [12] and de Blok et al. [23]. The cause of the discrepancy appears to be due to different choices of galaxy inclination when measuring the rotation curve (de Blok et al. [23]); NGC 6946 is an almost face-on galaxy ($32° - 38°$ inclination). Since another version of this rotation curve exists, measured by Boomsma [4], with values within 3% of Carignan et al. [12], we adopt the Carignan et al. value. Generally, inclinations are well defined, and situations like this are uncommon.

**Slope** The mean difference in $a_v$, the parameter that describes the slope at radii beyond the turnover radius, is 0.06 and the mode is 0.02. The value of this parameter is sensitive to the transcription of the published rotation curve values. We note that GalAPAGOS, for its family of solutions per target, permits a larger number of slope values as the turnover radius value increases. However, this does not have an effect on the difference values derived when comparing the "best fit" solution with the values from literature in Table 6. The largest difference (0.29) is for NGC 3621 which has a distorted outer disk. While this increases the uncertainty in the outer regions when using a tilted-ring model, it was even more of a challenge for GalAPAGOS to globally fit. (For this galaxy we used the kinematic parameter values derived from the rotation curve in the literature for the clustering analysis.)

data posed a problem for GalAPAGOS and the high inclination (76°) is a challenge for traditional rotation curve fitting packages.

In summary, differences in values derived for the kinematic parameters can be due to the adopted definition of $r_{out}$, distortions in the outer disk, the number of data points which trace the turnover radius region, inclination, data quality, and judgements made during the transcription of rotation curves from the literature. To characterize the amount of difference due to these factors we compare the subset of galaxies in our study that have GalAPAGOS rotation curves with fits of Eq. 1 to plots of their rotation curves as they appear in the literature (§ 3.1). The velocity amplitude differs between these two approaches by a few percent in the mean. The difference in the turnover radius in the mean is 0.07 and mode is 0.01 with a range from 0 to 0.18. For the slope the mean is 0.06, mode is 0.02 and the range is 0 to 0.29.

We adopt these values as uncertainties for our total sample in order to test the robustness of our kinematic classification scheme. For the velocity amplitude parameter we randomly vary the values in the total sample by up to 3% (its uncertainty). For the other two parameters we vary their values by adding or subtracting an uncertainty that is randomly selected from a normal distribution centred on the original value and spanning the range of difference values noted in the previous paragraph.

On the new set of values we re-run the hierarchical clustering routine described in § 3.2 and examine the 3D distribution in v—$r_{0,v}$—$a_v$ space. We iterate this approach numerous times, finding that the galaxies assigned to a particular class by the original clustering exercise tend to remain in a similar region of the parameter space. To show this we focus on the $a_v$—$r_{0,v}$ plane, since velocity differences are small, and overlay 6 of these runs in Fig. 11, where the left-hand plot shows all classes and right-hand plot only shows classes AS and AT. As an example, the right-hand plot figure clarifies that the majority of the galaxies originally assigned to the AT class, i.e. those with higher turnover radii, reside in a region unpopulated with AS galaxies in spite of the uncertainties.

The impact of uncertainties on the classification scheme is that it extends the range of the values per kinematic parameter but the adopted classes remain distinguishable.

Since the distribution of values appear to form a continuous 3-D shape in the v—$r_{0,v}$—$a_v$ volume, the boundaries between classes are not expected to be clearly delineated without establishing a convention for the demarkation. For the following investigations we continue to adopt the divisions implied by the original dendrogram associated with the hierarchical clustering analysis plus visual inspection of the 3-D parameter space (see § 4.1).

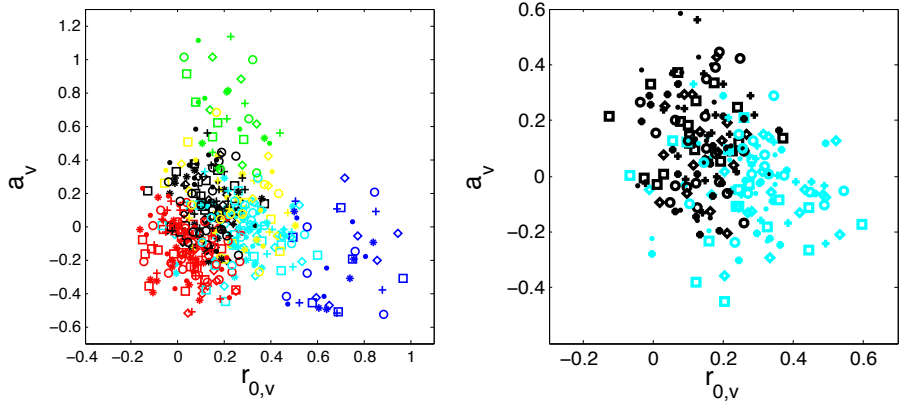

Figure 11: Impact of uncertainties on the $a_v$—$r_{0,v}$ plane for the total sample (left) and AS and AT classes only (right). Values for all three kinematic parameters were varied to represent the uncertainty as described in § 4.2 and projected onto the $a_v$—$r_{0,v}$ plane. Six variation are presented here; each symbol represents an iteration of the hierarchical clustering algorithm. The colours correspond to the different kinematic class designations: red = A, black = AS, cyan = AT, blue = S, and green = T. Yellow denotes non-classified outliers. Note that the axes span different ranges in the two plots.

## 5. Survey of ostensible correlations

A number of galaxy characteristics were explored in order to find or confirm correlations like those mentioned in the introduction. For the bulk of the galaxies, these include: absolute B magnitude, luminosity class, size of optical disk $D_{25}$, morphological type, presence of bar, HI content and mass ratio of gas mass over dynamical mass. Additionally, the mass model analysis of ten galaxies in Wiegert [81] provides the optical disk scale lengths used in this paper. Table 9 lists all galaxies according to their kinematic class, along with the values used to investigate the correlations. The following sections describe these values and the correlations in more detail. A summary is provided in § 5.8.

A number of the correlation investigations require the distance to the galaxy, both for converting the galaxy radius to kpc as well as for determining the HI mass and star formation rates. The distance is the source of the largest uncertainty in these calculations. In Table 9 we use Tully-Fisher determined distances for the galaxies from García-Ruiz et al. [31], which they deemed to be more accurate than Virgocentric inflow model distances. Still, García-Ruiz et al. [31] estimate an uncertainty of up to 30% for these distances.

*Note that in this section there are several interactive s2plot figures in the electronic pdf version of this paper, viewable with e.g. Acrobat Reader.* We have used the s2plot library described in Barnes et al. [2].

Table 9

| Galaxy | Class | D | ref | T | Bar | $D_{25}$ | $M_B$ | LC | SFR | $M_{HI}$ | $M_{dyn}$ | $M_{ratio}$ |
| --- | --- | --- | --- | --- | --- | --- | --- | --- | --- | --- | --- | --- |
| | (2) | (3) | (4) | (5) | (6) | (7) | (8) | (9) | (10) | (11) | (12) | (13) |
| NGC0801 | A | 80 | a | 5.3 | | 64.2 | -22.0 | | 2.46 | 1.94 | 64.5 | 0.042 |
| NGC0891 | A | 9.5 | b | 3 | | 36.0 | -20.2 | 5 | 0.64 | 0.27 | 20.7 | 0.019 |
| NGC2613 | A | 25.9 | c | 3.2 | | 57.4 | -21.8 | 3 | 0.61 | 0.91 | 72.3 | 0.018 |
| NGC2683 | A | 5.1 | d | 3 | | 14.1 | -21.1 | 4 | 0.04 | 0.03 | 9.7 | 0.005 |
| NGC2841 | A | 14.1 | e | 3 | | 28.3 | -20.8 | 1 | 0.19 | 0.57 | 94.2 | 0.008 |
| NGC2903 | A | 8.9 | f | 4 | B | 30.9 | -20.9 | 2 | 1.04 | 0.27 | 21.5 | 0.018 |
| NGC2998 | A | 67 | a | 5.2 | B | 47.6 | -21.7 | 2.3 | 2.59 | 1.6 | 42.4 | 0.053 |
| NGC3351 | A | 9.33 | g | 3.1 | B | 19.6 | -20.2 | 3 | 0.11 | 0.07 | 12.0 | 0.009 |
| NGC3521 | A | 8.5 | h | 4 | B | 20.6 | -21.1 | 3 | 0.47 | 0.28 | 33.0 | 0.012 |
| NGC3992 | A | 15.5 | i | 4 | B | 36.4 | -21.2 | 1 | | 0.3 | 38.8 | 0.011 |
| NGC4013 | A | 15.5 | i | 3 | | 22.0 | -19.4 | 5 | 0.27 | 0.14 | 18.0 | 0.011 |
| NGC4100 | A | 15.5 | i | 4.1 | | 20.6 | -20.6 | 2 | 0.35 | 0.18 | 11.7 | 0.022 |
| NGC4138 | A | 15.5 | i | -0.9* | | 13.2 | -18.7 | | 0.14 | 0.08 | 8.4 | 0.014 |
| NGC4157 | A | 15.5 | i | 3.3 | B | 28.1 | -19.8 | 3 | 1.31 | 0.35 | 20.3 | 0.024 |
| NGC4217 | A | 15.5 | i | 3 | | 24.5 | -20.0 | 5 | 0.85 | 0.12 | 10.6 | 0.016 |
| NGC4258 | A | 7.73 | g | 4 | B | 38.2 | -21.1 | 3.9 | 0.42 | 0.41 | 24.9 | 0.023 |
| NGC5033 | A | 11.9 | j | 5.1 | | 34.1 | -20.9 | 2 | 0.51 | 0.38 | 30.9 | 0.017 |
| NGC5055 | A | 7.2 | k | 4 | | 24.8 | -21.2 | 3 | 0.40 | 0.26 | 22.0 | 0.017 |
| NGC5371 | A | 34 | j | 4 | B | 39.4 | -22.1 | 1 | 1.42 | 0.75 | 41.5 | 0.025 |
| NGC5533 | A | 54 | a | 2.4 | | 45.0 | -21.6 | 1.9 | 1.03 | 1.88 | 87.7 | 0.030 |
| NGC5907 | A | 11 | l | 5.4 | | 36.2 | -21.0 | 3 | 0.19 | 0.42 | 34.1 | 0.017 |
| NGC6674 | A | 49 | a | 3 | B | 57.1 | -21.6 | | | 2.25 | 94.6 | 0.033 |
| NGC7331 | A | 14.52 | g | 3.9 | | 39.2 | -21.6 | 2 | 1.36 | 0.67 | 4.1 | 0.233 |
| NGC0300 | AS | 2.15 | m | 6.9 | | 12.1 | -19.2 | 6 | | 0.22 | 2.5 | 0.123 |
| NGC1003 | AS | 11.8 | n | 6 | | 11.9 | -19.5 | 5 | 0.05 | 0.35 | 9.5 | 0.052 |
| NGC2403 | AS | 3.18 | o | 6 | B | 18.5 | -19.7 | 5 | 0.02 | 0.18 | 9.9 | 0.025 |
| NGC2915 | AS | 5.6 | p | 1.6* | | 3.0 | | 9 | | 0.07 | 3.1 | 0.033 |
| NGC3198 | AS | 13.8 | g | 5.2 | B | 25.9 | -20.5 | 3 | 0.13 | 0.56 | 18.6 | 0.042 |
| NGC3600 | AS | 11.3 | q | 1* | | 6.2 | -17.8 | | 0.02 | 0.1 | 2.8 | 0.050 |
| NGC3621 | AS | 6.6 | g | 6.9 | B | 18.8 | -20.5 | 5.8 | 0.14 | 0.4 | 15.3 | 0.037 |
| NGC3726 | AS | 15.5 | i | 5.1 | B | 23.8 | -20.6 | 2 | 0.18 | 0.33 | 17.1 | 0.027 |
| NGC4096 | AS | 10.2 | q | 5.3 | B | 16.8 | -20.3 | 4 | 0.16 | 0.11 | 4.6 | 0.034 |
| NGC4183 | AS | 15.5 | i | 5.9 | | 19.3 | -19.7 | 5.9 | 0.02 | 0.19 | 5.4 | 0.050 |
| NGC5585 | AS | 7.6 | a | 6.9 | B | 9.4 | -18.5 | 7.1 | | 0.11 | 2.2 | 0.072 |
| NGC6503 | AS | — | | 5.9 | | 9.0 | -18.6 | 4.8 | 0.03 | 0.05 | | |
| NGC6946 | AS | 10.1 | r | 5.9 | B | 19.6 | -20.9 | 1.3 | 0.19 | 0.74 | 17.7 | 0.059 |
| UGC0128 | AS | 56.4 | a | 8.1 | | 31.2 | | | | 0.57 | 15.7 | 0.051 |
| UGC2259 | AS | — | | 7.8 | B | 4.7 | -16.0 | 6 | | 0.04 | | |

Continued on next page...

| Galaxy | Class | D | ref | T | Bar | $D_{25}$ | $M_B$ | LC | SFR | $M_{HI}$ | $M_{dyn}$ | $M_{ratio}$ |
| | (2) | (3) | (4) | (5) | (6) | (7) | (8) | (9) | (10) | (11) | (12) | (13) |
|---|---|---|---|---|---|---|---|---|---|---|---|---|
| UGC2459 | AS | 36.3 | q | 7.8 | | 24.6 | | | 0.11 | 0.8 | 22.4 | 0.050 |
| UGC3137 | AS | 33.8 | q | 4.3 | | 37.4 | -17.6 | 6 | | 0.84 | 15.1 | 0.078 |
| UGC6446 | AS | 15.5 | i | 6.6 | | 6.4 | -17.2 | 5 | | 0.16 | 2.0 | 0.114 |
| UGC6917 | AS | 15.5 | i | 8.8 | B | 11.3 | -18.5 | 5 | | 0.11 | 2.6 | 0.060 |
| UGC6930 | AS | 15.5 | i | 6.6 | B | 6.4 | -18.3 | 5 | | 0.15 | 3.9 | 0.056 |
| UGC6983 | AS | 15.5 | i | 5.8 | B | 8.6 | -18.4 | 5 | | 0.14 | 3.8 | 0.054 |
| NGC0055 | AT | 1.6 | s | 8.8 | B | 13.9 | -17.8 | 5.5 | | 0.13 | 1.7 | 0.109 |
| NGC0925 | AT | 9.12 | g | 7 | B | 28.4 | -20.1 | 4 | 0.02 | 0.35 | 0.8 | 0.588 |
| NGC3118 | AT | 21.7 | q | 4.1 | | 13.1 | -18.4 | | 0.06 | 0.21 | 2.4 | 0.120 |
| NGC3510 | AT | 8.8 | q | 8.6 | B | 4.0 | -17.6 | 4 | 0.01 | 0.06 | 0.8 | 0.104 |
| NGC3917 | AT | 15.5 | i | 5.8 | | 20.8 | -19.9 | 5.9 | | 0.09 | 5.3 | 0.025 |
| NGC3972 | AT | 15.5 | i | 4 | B | 16.6 | -18.9 | | 0.04 | 0.06 | 3.0 | 0.027 |
| NGC4010 | AT | 15.5 | i | 6.9 | B | 13.8 | -19.1 | | 0.14 | 0.15 | 3.1 | 0.070 |
| NGC4144 | AT | 6 | q | 6 | B | 9.1 | | 5 | 0.01 | 0.03 | 0.6 | 0.075 |
| NGC5023 | AT | 8 | q | 6 | | 14.5 | | | | 0.06 | 1.7 | 0.052 |
| NGC5229 | AT | 13.2 | q | 6.8 | B | 8.6 | | | 0.05 | 0.08 | 0.7 | 0.156 |
| NGC7793 | AT | 3.9 | t | 7.4 | | 11.8 | -18.9 | 6.3 | 0.01 | 0.06 | 3.8 | 0.021 |
| UGC3909 | AT | 24.5 | q | 5.8 | B | 11.6 | -17.1 | | | 0.19 | 1.9 | 0.139 |
| UGC5459 | AT | 15.9 | q | 5.2 | B | 17.5 | -19.4 | | 0.11 | 0.2 | 6.8 | 0.040 |
| UGC6399 | AT | 15.5 | i | 8.8 | | 9.7 | -17.7 | | | 0.05 | 1.2 | 0.052 |
| UGC7321 | AT | 14.9 | q | 6.6 | | 20.9 | | | | 0.15 | 3.2 | 0.066 |
| UGC7774 | AT | 20.6 | q | 6.3 | | 12.0 | -16.5 | | | 0.17 | 2.5 | 0.096 |
| M33 | S | 0.84 | u | 5.9 | | 15.2 | -19.4 | 4 | | 0.2 | 2.8 | 0.102 |
| NGC0247 | S | 2.8 | a | 6.9 | B | 16.0 | -18.6 | 6.9 | 0.0008 | 0.08 | 2.9 | 0.039 |
| NGC1560 | S | — | | 7 | | 8.3 | | | 0.0013 | 0.03 | | |
| UGC7089 | S | 11.6 | q | 7.9 | | 10.3 | -17.8 | | | 0.04 | 0.9 | 0.068 |
| UGC9242 | S | 12.6 | q | 6.6 | | 15.4 | -19.8 | | | 0.05 | 2.3 | 0.031 |
| NGC2770 | T | 21 | q | 5.3 | | 21.1 | -20.7 | | 0.20 | 0.26 | 6.8 | 0.053 |
| NGC2976 | T | 3.6 | v | 5.2 | | 6.0 | -18.1 | 7 | 0.01 | 0.01 | 0.4 | 0.034 |
| NGC3556 | T | 11.6 | x | 6 | B | 13.4 | -20.7 | 5.9 | 0.86 | 0.34 | 9.2 | 0.053 |
| NGC4389 | T | 15.5 | i | 4.1 | B | 11.3 | -18.4 | 7 | 0.14 | 0.04 | 1.3 | 0.038 |
| UGC1281 | T | 5.1 | q | 7.5 | | 7.7 | | | | 0.02 | 0.3 | 0.072 |
| UGC6818 | T | 15.5 | i | — | B | 7.9 | -16.3 | | | 0.06 | 0.8 | 0.114 |
| NGC3877 | O | 15.5 | i | 5.1 | | 24.2 | -20.4 | 4 | 0.29 | 0.09 | 6.4 | 0.020 |
| NGC3949 | O | 15.5 | i | 4 | | 10.2 | -19.9 | 5.9 | 0.84 | 0.17 | 4.1 | 0.057 |
| NGC3953 | O | 15.5 | i | 4 | B | 27.7 | -21.4 | 1.1 | 0.15 | 0.17 | 14.5 | 0.016 |
| NGC4051 | O | 15.5 | i | 4 | B | 22.0 | -20.0 | 3 | 0.67 | 0.16 | 5.8 | 0.038 |
| NGC5301 | O | 22.5 | q | 4.7 | B | 25.9 | -19.8 | 4 | 0.24 | 0.3 | 7.1 | 0.059 |
| UGC2885 | O | 79 | a | 5.2 | | 98.9 | -22.0 | | 1.92 | 3.38 | 149.6 | 0.032 |

| Galaxy | Class | D | ref | T | Bar | $D_{25}$ | $M_B$ | LC | SFR | $M_{HI}$ | $M_{dyn}$ | $M_{ratio}$ |
| --- | --- | --- | --- | --- | --- | --- | --- | --- | --- | --- | --- | --- |
| | (2) | (3) | (4) | (5) | (6) | (7) | (8) | (9) | (10) | (11) | (12) | (13) |
| UGC6667 | O | 15.5 | i | 5.9 | | 5.6 | -17.8 | | | 0.05 | 1.2 | 0.059 |
| UGC8246 | O | 19.4 | q | 5.9 | B | 14.0 | -17.1 | | | 0.12 | 0.9 | 0.196 |

Table 9: The total sample of galaxies listed in order of their classes, with values for the galaxy characteristics used in the subsequent correlation analysis. (2) Kinematic class; (3) Distance (Mpc); (4) Distance references: a) Sanders [67], b) van der Kruit [77], c) Irwin and Chaves [36], d) Casertano and van Gorkom [16], e) Macri et al. [48], f) Karachentsev et al. [41], g) Freedman et al. [30], h) Zeilinger et al. [84], i) Sanders and Verheijen [68], j) Begeman [3], k) Pierce [59], l) Barnaby and Thronson [1], m) Karachentsev and Kashibadze [42], n) Broeils and Knapen [9], o) Madore and Freedman [49], p) Meurer et al. [53], q) García-Ruiz et al. [31], r) Rogstad and Shostak [64], s) Puche et al. [61], t) Karachentsev [39], u) Madore and Freedman [49], v) Karachentsev et al. [40], x) King and Irwin [45] ; (5) de Vaucouleurs revised morphological type from the second catalogue of bright galaxies (RC2). Note "—" means type is unavailable, "*" means the RC2 type differs from other sources; (6) Presence of bar; (7) optical diameter of galaxy measured out to the 25 mag/$''^2$ isophote [kpc]; (8) absolute B magnitude; (9) Luminosity class; (10) Star formation rate of stars with masses $\geq 5 M_\odot [M_\odot/yr]$; (11) Derived HI mass [$\times 10^{10} M_\odot$]; (12) Derived dynamical mass [$\times 10^{10} M_\odot$]; (13) Ratio of gas mass (HI mass multiplied by 1.4 to include helium) over dynamical mass. Values in columns (5), (6), (7), (8) and (9) are from the extragalactic database HyperLeda (Paturel et al. [57]). The derivation of the values in column (8) is described in § 5.6.

*5.1. Absolute B magnitude*

The first investigation serves as a 'sanity check' of the rotation curve data – all literature, including of course the Tully-Fisher relation, agree on the correlation between the rotational velocity of a galaxy and its luminosity, in which more luminous galaxies also have high rotational velocities (e.g. [71]). Using B magnitude values from the extragalactic database HyperLeda (Paturel et al. [57]), we find a clear correlation between maximum rotational velocity and absolute B magnitude, in that the fast rotating A class galaxies are also the brightest. However no correlation can be discerned between brightness and the lower velocity S and T branches, since the few galaxies in these categories (especially T) show a wide range of magnitudes.

This correlation with velocity is also present for the luminosity class and the kinematic classes other than S and T. Luminosity class is a classification based on galaxy luminosity in combination with the degree of development of spiral arms, a classification that was introduced by van den Bergh [74]. The values listed in Table 9 are on a scale from 1-11, where 1 = supergiant galaxy, 3 = bright giant galaxy, 5 = normal giant galaxy, 7 = subgiant galaxy , 9 = dwarf galaxy.

*5.2. Morphology*

We investigate the potential correlation of the morphological type (Hubble type) and the rotation curve parameters. According to Noordermeer et al. [55] there is disagreement on whether there is a correlation between morphology and rotation curves. However, their analysis indicates that Sa, Sb, and Sc galaxies show a correlation with decreasing maximum rotational velocity [65], although they did not find any correlation with any other rotation curve properties.

The morphological type has been expressed in numbers in the de Vaucouleurs system described in the Second Reference Catalog of Bright Galaxies (RC2) and the values in column (5) of Table 9 are from the HyperLeda version of these values. The galaxies in our sample range from -0.9 to 8.8, where -0.9 corresponds to early type S0, through Sa (1), Sab (2), Sb (3), Sbc (4) etc to later types Sc (5), Sd(7) and Sm (9, magellanic type). We note that the galaxies designated 1 and below have disks but their morphogical types in the RC2 differ from types reported in other sources, as described below.

Fig. 12 shows the morphology displayed as colours where bright colours correspond to earlier type galaxies (low values) and darker colours correspond to later type galaxies (higher values). Fig. 13, which shows morphology as a function of kinematic class, further illustrates the results. While a trend is apparent in which earlier morphological type corresponds to higher rotational velocity, it is not a particularly tight correlation. Additionally it is difficult to draw conclusions about trends of morphology with either the turnover radius or the slope, as one can see by rotating the figure to view the $r_{v,0}$—$a_v$ plane. However we note that the majority of the A class galaxies are Sb and Sbc. The one extremely early type galaxy, NGC 4138, is probably an Sa [73], which changes its type from -0.9 to 1. The range for the sparsely populated T class includes Sbc to Sd galaxies. While the AS, AT and S classes span a similar range, Scd and Sd galaxies are typical.

Apparent exceptions in the AS class are NGC 3600, with NED designation Sa? (i.e. "doubtful"), and NGC 2915, with NED designation I0 (i.e. an S-shaped irregular). Although these discrepancies bring into question their de Vaucouleurs designations, we include these galaxies' HyperLeda types when considering trends, summarized in § 5.8.

*5.3. Size of optical disk, $D_{25}$, and scale length*

A strong correlation between the size of the optical disk and rotation curve is found by Chattopadhyay and Chattopadhyay [17]. $D_{25}$ is defined as the length of the projected major axis at the isophotal level 25 mag/arcsec$^2$ measured in the B-band. We retrieved the $D_{25}$ values from the HyperLeda database, measured as angular sizes for all galaxies in the sample. These values have not been corrected for any extinction effects or inclination [56]. Using the galaxy distances listed in Table 9, the diameters were converted to units of kpc, and their values are visualized using the colours of the data points in Fig. 14.

Figure 12: The parameter distribution in three dimensions (vertical axis = $v_0$, horizontal axes = $r_{0,v}$ (side) and $a_v$ (front)) with the data points coloured by morphology to visualize correlations with galaxy characteristics. Light colours correspond to early type galaxies, darker colours correspond to late type galaxies.

We confirm the strong trend in which increases in maximum velocity occur with increasing $D_{25}$. A synopsis of mean radii per class is provided in Table 10 in the summary in § 5.8. However, perhaps a more interesting parameter to investigate is the scale length. We have measured the exponential disk scale length for 10 galaxies, 7 of which are Class A, from luminosity profiles constructed from SDSS I-band data (§ 2.2). Although we can not investigate potential differences that may exist between the kinematic classes due to the low number, a correlation would be suggested if a particular small range of values for an A class parameter mapped to a small range of values for disk scale length. In Fig. 15, we investigate whether such a correlation is implied between disk scale length and maximum velocity, outer slope of the rotation curve, and turnover radius.

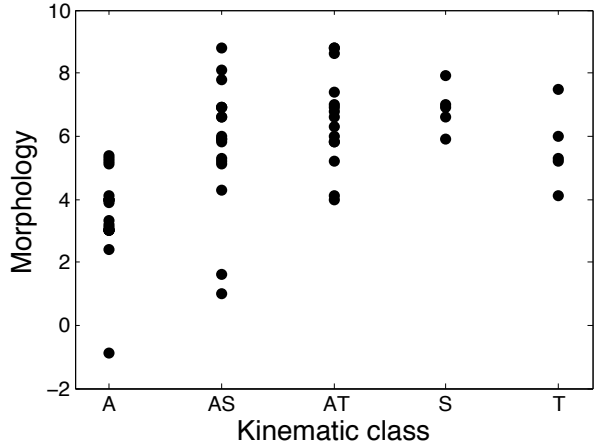

Figure 13: Kinematic class plotted against morphological type in the de Vaucouleurs RC2 system. Note that the three earliest type galaxies in this plot have more than one reported Hubble type designation, as described in § 5.2.

In contrast to the results of Chattopadhyay and Chattopadhyay [17], who found a weak correlation between disk scale length and the shape of the rotation curve, we find no hint of such a correlation. That is, for the narrow range of values associated with the A class, for each parameter describing our fitted rotation curves, there is a large range of disk scale length values. However not only could our finding be due to our limited sample size but also due to difference in the way we characterize the rotation curve shape compared to the way Chattopadhyay and Chattopadhyay [17] characterize the rotation curve.

*5.4. Bar*

Despite the indication of a weak correlation between the presence of a bar and rotation curve shape found by Chattopadhyay and Chattopadhyay [17], there was no such correlation noticed for these galaxies, as can be discerned directly from Table 9. Note however that we do not examine the inner part of the rotation curve (its inner slope) in detail. The inner part is here only determined by the turnover radius parameter. Also, the possible 'hump' at the turnover, that can indicate the existence of a bar, is not visible in the rotation curves of our samples due to low resolution (few data points). Therefore, while the classes do not appear to be sensitive to the effect investigated by Chattopadhyay and Chattopadhyay [17], this is not conclusive.

*5.5. HI content, dynamical mass and mass ratio*

Since mass is the main ingredient of a galaxy's kinematic behaviour, this is one of the most interesting correlations to study. Unable to compare the DM content for these galaxies in a homogenous way, we instead take a look at the dynamical mass calculated at the apparent edge of the HI disk ($r_{out}$), which is an indication of the lower limit of the total (including DM) mass content. Additionally, we calculate the HI content, and investigate the mass ratios in the different kinematic classes.

Column (12) in Table 9 is the dynamical mass of the galaxy at $r_{out}$, the outer radius measured for the rotation curves of these galaxies. It is derived by the simple relation

$$M(r) = \frac{v(r)^2 r}{G} \qquad (2)$$

Figure 14: The colour of the data points are scaled to the value of the size of the optical disk, $D_{25}$. Dark colours refer to a large disk diameter.

where G is the gravitational constant and v is measured at $r = r_{out}$. The three rotation curves in Hoekstra et al. [34] are normalised and they do not report values for $r_{out}$, so are excluded.

The HI content (Column 11 in Table 9) was derived for the remaining 76 galaxies by taking the integrated HI flux, as given in magnitudes in the de Vaucouleurs et al. [25] RC3 catalogue, and converting it from magnitudes to the integrated flux following equation 3 (from RC3):

$$m_{HI} = -2.5 \log S + 17.40, \tag{3}$$

where the integrated flux $S$ is in units of Jy km/s. The flux was then converted to mass in units of $M_\odot$ by the standard formula (eq. 4):

$$M_{HI} = 2.36 \times 10^5 D^2 S, \tag{4}$$

where D is the distance to the galaxy in Mpc. For a few galaxies, HI masses are available for comparison [see e.g. 67, 68]. Generally, our values are lower, with a mean difference of $2.6 \times 10^9 M_\odot$, or 32%.

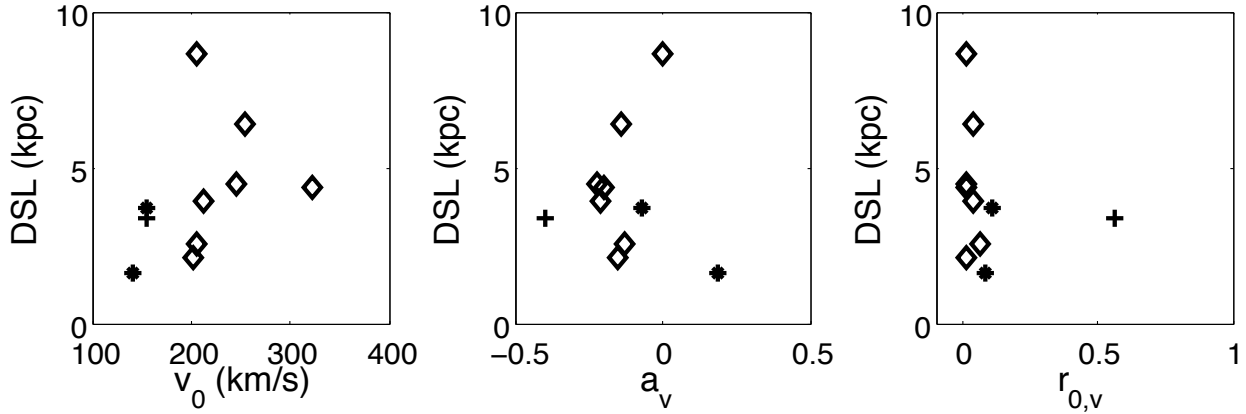

Figure 15: 10 mass modelled galaxies: Rotational velocity ($v_0$) plotted against the disk scale length (left), outer slope of the rotation curve ($a_v$) plotted against the disk scale length (middle), and the turnover radius ($r_{0,v}$) plotted on the right. Data points displayed as diamonds are A class galaxies, asterisks are AS and the T class galaxy is shown with a '+'. Note that the uncertainties for the disk scale length in this figure (although not shown) are at least 30%, mainly as a consequence of the distance to the galaxy which is used to convert the scale from units of arcseconds to kpc.

Assuming the chemical composition is similar in other galaxies to the solar vicinity, the HI mass is scaled by 1.4 to account for helium [e.g. 35, p.78]. The gas fraction $M_{gas}/M_{dyn}$ is listed in Column (13) of Table 9. Fig. 16 and Fig. 17 show the HI mass and this gas fraction, respectively.

Although we do not show it in a figure, there is a correlation between increasing dynamical mass and increasing velocity (i.e. velocity dominates over radius in Equation 2). While a strong trend does not exist, it is notable that the galaxies with the most amount of HI gas (in units of mass) are A class galaxies.

Although there is a scatter in values for the A class, this class has a median $M_{gas}/M_{dyn}$ mass ratio which is three times lower than the other classes with a value of 2%; the other classes each have a median mass ratio near 6%. However, it is likely that the dynamical mass in the ratio dominates over the (probably underestimated) HI mass, and the correlation seen in Fig. 17 is simply an expression of the velocity component in the dynamical mass equation.

*5.6. Star formation rate*

Radio continuum emission at 20 cm wavelength can be used as a tracer of the star formation rate (SFR) in a galaxy. Condon [19] reviews how the emission is produced by supernova remnants from massive (more than $8M_\odot$) and short lived (less than $3 \times 10^7$ years) stars. This emission thus traces recent star formation activity in the galaxy. An additional advantage of using this tracer is that the observed flux densities are proportional to intrinsic luminosities since radio wavelengths suffer minimal extinction.

Integrated flux from 20 cm continuum data from the NRAO VLA Sky Survey (NVSS) [20] is available for 55 of the 79 galaxies used for our explorations. Since NVSS only can properly image structures of angular size 10′ or less, larger galaxies in our sample might be missing a significant fraction of the flux. This is an issue for eight galaxies in the sample which have an optical size ($D_{25}$) between 10′ and 20′. Two of these 8 galaxies have such large angular sizes that we exclude them from the SFR analysis altogether (NGC 55 and M33).

The flux for the remaining 53 galaxies is translated to luminosity (for a particular frequency, in this case 1.4 GHz)

Figure 16: The colour of the data points are scaled to the value of the HI mass. Light colours refer to larger amounts of mass.

using the inverse square law:

$$L_\nu = 4\pi D^2 \times Flux_\nu \qquad (5)$$

Equation 21 in Condon [19] was used to derive star formation rate, following:

$$SFR = \frac{L_\nu}{5.3 \cdot 10^{21} \cdot \nu^{-\alpha}} \qquad (6)$$

where SFR refers to formation of stars with masses $\geq 5$ $M_{sun}$ in units of $M_{sun}$/year, $\nu$ is the frequency of emission in GHz, and $\alpha$ is the radio spectral index. We use $\alpha = 0.8$, typical for galaxies at GHz frequencies [e.g. 19, 54]. The resulting star formation rates are listed in Table 9.

Here we investigate the relationship between SFR and gas fraction as a function of kinematic class. Fig. 18 shows the classes vs SFR, where the datapoints are coloured by gas fraction ($M_{gas}/M_{dyn}$) such that lighter colours are associated with smaller gas-to-total-mass ratios. The shapes of the symbols represent the morphological types; there are only five dwarf galaxies and they reside in AT, S and T classes.

Figure 17: The colour of the data points are scaled to the value of the gas fraction ($M_{gas}/M_{dyn}$. Light colours refer to a higher gas fraction.

In Fig. 18 we can see that the A class contains the galaxies with the highest SFR, although only those with high gas fraction exhibit high SFR rates. The figure also shows that classes AS and AT, which have high gas fractions and Sd as well as Sc galaxies, do not have high SFR. This quiescence is nevertheless expected from other studies which show that a galaxy's star forming phase depends on factors such as gas density at various spatial scales [see 43, for a review] or exhibits trends with rotational velocity regimes [e.g. 47].

When plotting the kinematic parameters vs. star formation rate, we find a strong indication of a correlation between velocity amplitude $v_0$ and SFR. Fig. 19 shows this correlation. The S and T classes are not well represented, making it hard to distinguish any correlation with the slope or turnover radius.

It is expected that there is a relationship between high rotational velocity and SFR, due to the consequential higher angular velocity of density waves which creates conditions favourable for star formation, as mentioned by, for example, Zasov and Smirnova [83]. Observationally the seven Sc galaxies in this sample that show this trend are rather quiescent – that is their SFRs are less than or equal to the Milky Way's rate of 2-10 $M_\odot$/yr. Additionally the velocities for galaxies in class A with the highest star forming rates ( > 1 $M_\odot$/yr) are $\gtrsim$ 200 km/s. It is probably because only 5/53

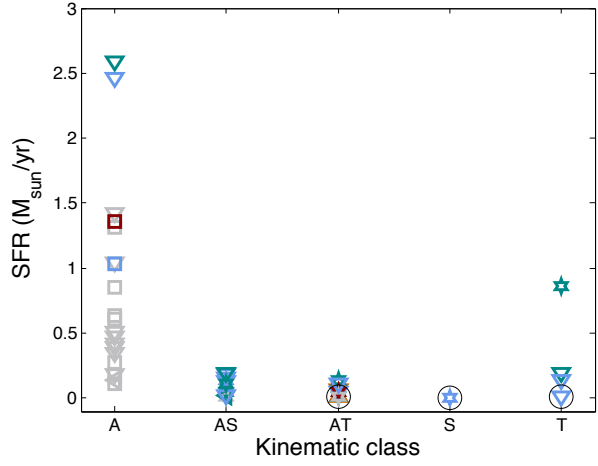

Figure 18: Relation between kinematic class, SFR and gas fraction ($M_{gas}/M_{dyn}$). The shades of the data points, from light to dark, represent the gas fraction listed in Table 9: light grey represents low gas-to-total-mass ratios while dark red represents the high gas mass ratio. The symbols represent the different morphological types: ◁= Sa (de Vaucouleurs morphological type -1 to 2), squares= Sb (2–4), ▽= Sc (4–6), stars= Sd (6–8), and △=Sm/Irr (8–10). Black circles mark dwarf galaxies.

.

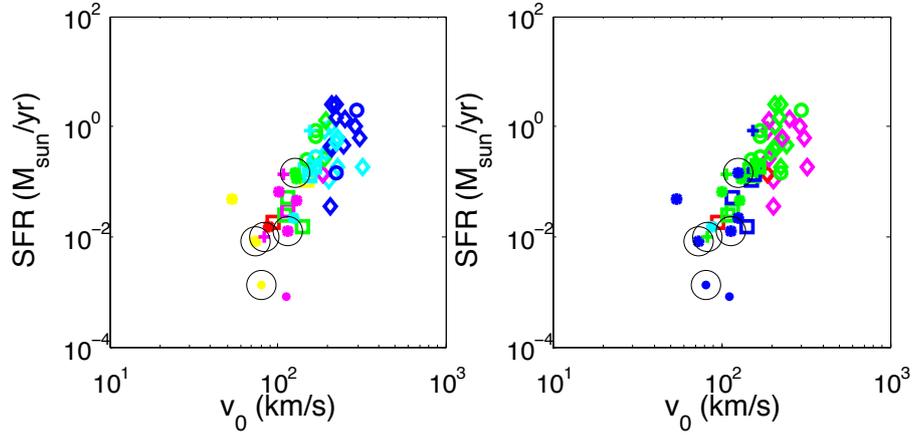

Figure 19: Star formation rates calculated for 53 of the galaxies in the sample are plotted against the velocity amplitude $v_0$ in this log-log plot. The left hand plot is coloured by magnitude: blue= -21, cyan=-20 to -21, green=-19 to -20, magenta= -18 to -19, red=-17 to -18. The right plot is coloured by morphological type: red=Sa (de Vaucouleurs morphological type -1 to 2), magenta=Sb (2–4), green=Sc (4–6), blue=Sd (6–8), cyan=Sm,Irr (8–10). The symbols denote the kinematic classifications given in this paper: diamonds=A, squares=AS, asterisk=AS, dots=S, '+'-signs=T, o=no class given (outliers). Black circles mark dwarf galaxies.

galaxies in our sample are dwarfs that this trend is in contrast with the tendency of decreasing SFR with increasing velocity found by Lee et al. [47] in their dwarf dominated sample of galaxies with rotational velocities larger than 120 km/s and brighter than $M_B$ = -19.

In addition, since we know that the 1D velocity dispersion plays a role in star formation laws [e.g. 51] such as the Kennicutt-Schmidt law [44], it would be illuminating to complement these data with 1D velocity dispersion data as well. GalAPAGOS constrains this 1D velocity dispersion as one of its parameters, and thus we have this information for 15 of the galaxies in the sample of Wiegert [81]. It is however hard to draw any conclusions with these few datapoints, as shown in Fig. 20. Only a possible constraint can be seen ($v_\sigma$ between 10 and 20 km/s), but there are no visible trends between velocity dispersion and SFR.

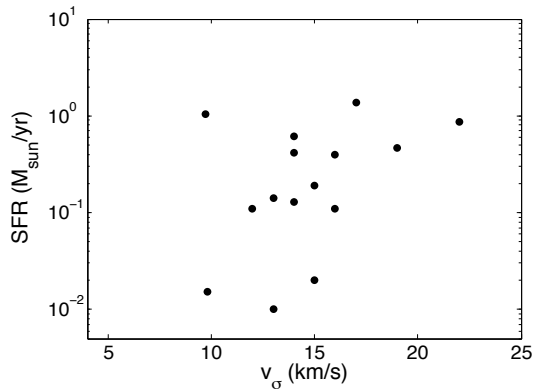

Figure 20: Star formation rates are plotted against the 1D velocity dispersion for 15 galaxies.

*5.7. Slope parameter and declining rotation curves.*

Although $a_v$ values $\leq$ -0.2 imply that the slope of the fit in Eq. 1 is decreasing, this can be too gentle to be significant. Instead we use the definition for "declining" proposed by Corradi & Capaccioli (1990) which is that the velocity falls by $\gtrsim$ 20% from $v_0$, beyond the turnover radius. For Class A galaxies NGC 3521 and NGC 5055 (Table 8), we have derived rotation curves using both GalAPAGOS and GIPSY's ROTCUR task on THINGS data [23] allowing us to explore this in some detail. Given their potential decreases in velocity outlined below, we retain these as declining rotation curve candidates. Considered together, the ROTCUR and GalAPAGOS curves for NGC 3521 suggest the velocity could drop by as much as 22% at a radius of ~25 kpc.

NGC 3521 is quite a contentious galaxy in that Casertano & van Gorkom (1991) found a steep decline past this radius using less resolved data. However this is not replicated in the analysis of de Blok et al. (2008) although they find that the velocity of the rotation curve's outer region is lower than that of its inner region. Using the THINGS dataset and considering the extent of the uncertainties in both our ROTCUR plot and that of de Blok et al. (2008), we find the drop could be as little as 12% or as large as 31% at r~25 kpc.

For NGC 5055 the GalAPAGOS curve drops by 19% at r~32 kpc while the largest difference in velocity (20% drop) occurs at r~25 kpc in the ROTCUR curve and then the velocity may flatten out, extending to 30 kpc (17% drop). This differs slightly from the analysis by de Blok et al. (2008), which also uses ROTCUR but the resulting curve has a slower decrease in velocity. They may have fit the same value of inclination and same value of position angle for all the tilted rings in their model. However the inclination and position angle change value at about 420″, and we followed a suggestion by R. Swaters (private communication) to use one set of values for these parameters in the inner region and another set of values for the outer region. Nevertheless at an outer radius of 50 kpc their curve's velocity is lower by 19%; using the uncertainties in their plot the velocity difference between the inner and outer regions could be between 5% and 32%.

Other galaxies in the full sample with velocity amplitude decreases of $\gtrsim$20% are NGC 5533 (21%), NGC 2683 (26%) and NGC 4138 (20%). While NGC 5533 and NGC 4138 have inclinations of 60° and 65° respectively, NGC 2683 has an inclination of 90° which makes modelling both difficult and suspect.

*5.8. Summary of analysis*

Table 10 presents the *average* rotation curve characteristics, in *relative* terms, of our preliminary kinematic classification scheme, described in § 4.1, along with possible correlations between each class and general galaxy characteristics. This investigation was guided by correlations between rotation curve behaviour and a number of galaxy characteristics previously described in the literature.

We found indications of correlations, in particular between the maximum rotational velocity parameter and absolute B magnitude, morphology and size of optical disk, which are in agreement with literature. For example, there is gradual increase in optical diameter with increase in velocity and this gradient is reflected between classes that have differing velocities. Since class A has the largest velocity range along with the fastest rotating galaxies, the trend of increasing $M_B$ with increasing velocity is most apparent in this class. The morphological correlation is more subtle in that Class A is dominated by Sb and Sbc while in all other classes Scd galaxies prevail.

Sufficient data were available to explore relationships between HI gas content, dynamical mass, and SFR for a subset of 53 galaxies. While the other classes have some galaxies with higher gas fractions ($M_{gas}/M_{dyn}$), the galaxies with the largest HI content and highest SFR reside in Class A and have the latest Hubble Types within that class (Sc to Scd). Since much of this sub-sample resides in Class A we can clearly see the known correlation [e.g. 44] that the star formation rate generally increases with rotational velocity (Fig. 19).

Although we looked for correlations between the kinematic class and the presence of bar or optical scale length or 1-D velocity dispersion, we found none present. However our explorations for a relationship between the stellar disk and rotation, or an effect due to a bar, were limited in that we do not fit the inner part of the rotation curve in detail. Additionally, for the velocity dispersion and the optical scale length correlations we were limited by only having measurements for subsamples of 10-15 galaxies. Nevertheless the large range in values for the stellar disk scale lengths within Class A suggest that this optical property does not correlate with rotation curve characteristics.

| Properties | A | AS | AT | T | S |
|---|---|---|---|---|---|
| $v_0$ values | higher | intermediate to lower | lower | lower | lower |
| $r_{0,v}$ lengths | shorter | intermediate | intermediate | longer | intermediate |
| $a_v$ behaviour | flat or decreasing* | flat | flat | flat or decreasing | increasing |
| $M_B$ correlation | strong | intermediate | intermediate | none | none |
| Morphology | Sb/Sbc | Scd | Scd | Sd | Scd |
| $D_{25}$ [kpc] | 30 | 15 | 13 | 13 | 11 |
| HI [$\times 10^9 M_\odot$] | 5 | 3 | 1 | 1 | 1 |
| HI/$M_{dyn}$ (%) | 3 | 6 | 7 | 5 | 6 |
| SFR [$M_\odot$/year] | 0.7 | 0.1 | 0.05 | 0.001 | 0.4 |

Table 10: A summary in *relative* terms of the classes, their *average* characteristics and correlations with other galaxy properties. $M_B$ increases with increasing velocity and the 4th row indicates whether this known trend is discerned within each class. The other rows list the average or typical value of the correlation parameter per class. (*) Note that "decreasing" is not synonomous with "declining" (§ 5.7).

It is hard to discern conclusive correlations with the other two parameters, turnover radius ($r_{0,v}$) and outer slope ($a_v$), other than how they are already correlated with velocity. That is, high velocity Class A galaxies also have short turnover and low slope values and therefore galaxies with these three attributes are likely to appear to have correlations with velocity. At the other end of the velocity range, the S and T classes, the number of galaxies is low, so trends are not pronounced.

Finally, defining declining rotation curves as those that drop by about 20% or more after the turnover radius, we have four candidate galaxies, NGC 3521, NGC 4138, NGC 5055 and NGC 5533, all from Class A.

## 6. Discussion

A question that arises is whether the distribution of data points in the classification scheme (see Fig. 7) is possibly due to a selection bias. There is a 'void' in the centre between the two branches, where the galaxies exhibiting both large turnover radii in combination with strongly rising curves (high slope values) would exist. Galaxies with such rotation curves do exist; they are often irregular, dwarf or interacting galaxies. Indeed, some of the interacting galaxies that were excluded from our sample were pairs with such rotation curves, e.g. NGC 3109 and DDO170 from Hoekstra et al. [34]. There are at least ten dwarf galaxies in our sample, which reside in classes AS and AT as well as S and T, some bordering the void. A few of the outliers listed in Table 7 are also present in the void.

To answer the question above, it would be interesting to explore whether non-interacting galaxies avoid the void, while interacting ones are more commonly part of it. This needs however be confirmed by studying the rotation curves of a sample of galaxies confirmed to be interacting. More dwarf galaxies should be introduced into the analysis as well.

The distribution of the parameters suggests a relationship between maximum velocity and turnover radius plus slope of the outer rotation curve. That is, high velocity galaxies with particularly high values of turnover radius and slope are rare. Additionally, low velocity galaxies with short turnover radius and low slope values seem absent. Future studies will assess if this is an indication that the turnover radius and slope values are a direct physical consequence of the maximum velocity of the rotation curve, or due to a bias in the galaxy sample, or a consequence of our parameterization of the rotation curve.

Guided by previous studies, these preliminary classes have been examined for trends with observed galaxy characteristics. The average properties per class are listed in Table 10. The illuminating parameter is $v_0$ since Class A has the largest range of velocity amplitudes, and includes the highest velocities, while the velocity decreases through AS and the low velocity amplitude galaxies reside in AT, T, and S. Thus when examining the kinematic parameters for trends we found, in particular, correlations between higher maximum rotational velocity and the following observed properties: higher brightness, earlier type morphology, larger size of the disk ($D_{25}$) and higher star formation rate (as derived using radio continuum data). Our analysis also suggests that lower velocities are associated with a higher ratio of the HI mass over the dynamical mass. The correlations between investigated properties generally agree with those found in the literature. For example, a correlation between the morphological (Hubble) type and velocity amplitude is supported, which was not clear in early investigations using only optical rotation curves.

While we looked for correlations between the kinematic class and the presence of a bar or 1-D velocity dispersion, we found none present. However these assessments were probably hindered either by the limited number of points per rotation curve or by a small subsample size. Although our subsample of Class A galaxies with known stellar disk scale lengths is also small, the fact that the latter span a large range of values indicates that this optical property probably does not correlate with rotation curve characteristics.

The occurance of declining rotation curve candidates is a somewhat unexpected result (§ 5.7) since we specifically targeted galaxies we believed, at the beginning of this study, to be non-interacting [interactions, current or recent, would cause such rotation curve declines, see e.g. 80]. There might be other causes to this velocity behaviour. Decreasing velocity can be found in high surface brightness galaxies with short disk scale lengths [16] and grand design spiral arm galaxies [28]. Apart from NGC 4138 which has a shorter disk scale length of 1.2 kpc [38], we note that none of the other three Class A candidates in our sample conform to any of these groups. For example, they have intermediate or long disk scale lengths: NGC 3521 [4.5 kpc; 81], NGC 5055 [4 kpc; 81], and NGC 5533 [9.1 kpc; 55, whose rotation curve fit is also declining]

Additionally, the candidates do not have grand design patterns (NGC 3521; flocculent [e.g. 27], NGC 4138; flocculent in the outer regions [11], NGC 5055; flocculent [e.g. 27], NGC 5533; multi-tier disk [70]).

Upon revisiting the literature, there are indications that the galaxies with declining rotation curves in our sample are currently interacting, or have in the past, with minor mass companions. For example, NGC 3521 is part of sample of local galaxies each of which displays stellar tidal debris in the high sensitivity investigation by Martínez-Delgado et al. [50]. NGC 4138 is known to have two counter-rotating stellar disks [38], indicating an interaction history [32]. Chonis et al. [18] find that their photometry of an apparent tidal stream associated with NGC 5055 is consistent with the disruption of a companion galaxy. Additionally, van Eymeren et al. [78] place the lopsided galaxy NGC 5533 into a kinematic discrepancy category that is often associated with small companions.

We plan to investigate the effect of different parameterizations of the rotation curve on our classification scheme. However, in terms of more significant improvements, we are particularly looking forward to HI line synthesis projects on the upcoming Square Kilometre Array (SKA) precursor telescopes, the Australia SKA Pathfinder (ASKAP) and Karoo Array Telescope (MeerKAT) facilities. MHONGOOSE, on the latter, will collect high sensitivity 3D data on a few dozen galaxies while WALLABY, on the former, will spatially resolve ~1000 galaxies.

The homogeneity of the WALLABY survey, with results clarified by the MHONGOOSE examples, will allow us to reduce our uncertainties, particularly those associated with the outer radius. Apart from the rotational velocity which has a low uncertainty (3%), the outer radius measurements garnered from inhomogeneous data generate significant uncertainties in the other two kinematic parameters. WALLABY will enable more precise values for the turnover radius and outer slope by providing a consistently detected outer radius, perhaps specified at an adopted column density isophote. Also, although the general regions associated with our classes, within the $v_0$—$r_{0,v}$—$a_v$ volume, are robust, we have adopted boundaries between classes. While these were suggested by the hierarchical clustering results and visual inspection, we expect that applying this approach to the WALLABY data will provide improved information for selecting demarkations within the kinematic scheme.

Ancillary optical data will be available for these HI projects, for example, by the Australian National University's SkyMapper facility. Therefore, as well as clarifying the subjects described above, one will be able to mass model galaxies to assess their dark matter content and derive dark matter halo parameters. Incorporation of these parameters into our classification scheme could illuminate the strength of the role played by the gravitational potential in galaxy evolution and star formation efficiency.

# 7. Conclusions

We have explored the possibility of creating a classification system for galaxies based on their kinematics. Our kinematic classification scheme, based on HI rotation curves from the literature, is still in its first, exploratory, form. It was arrived at by applying to 79 disk galaxies the simple rotation curve model in Eq. 1, parameterized using the maximum rotational velocity $v_0$, turnover radius $r_{0,v}$ and outer slope $a_v$. We used hierarchical clustering as well as visual inspection to segment the 3D structure in the $v_0$—$r_{0,v}$—$a_v$ volume (e. g. Fig. 7) into the five classes described in § 4.1.

There are indications of correlations between these classes and observed galaxy characteristics of the galaxy (see Table 10). In particular trends with the $v_0$ parameter are conspicuous, such as higher maximum rotational velocity and the following observables: higher brightness, earlier type morphology, larger size of the disk ($D_{25}$) and higher star formation rate (derived using radio continuum data). We find a suggestion that lower velocities are associated with a higher ratio of the HI mass over the dynamical mass. Additionally, we find support for a correlation between the morphological (Hubble) type and velocity amplitude. This reproduction of some known correlations provides encouragement for continuing to improve upon the delineation of the kinematic classes.

Significant improvements of our uncertainties will be found using larger and homogenous data sets, such as those produced by HI line synthesis projects, on the SKA precursor telescopes, MHONGOOSE and WALLABY. The inclusion of more dwarf galaxies and interacting galaxies can then be tested, in order to observe the robustness of the 3D structure and how it could possibly change. That is, could interacting galaxies mainly occupy the central 'void' of the distribution?

The investigation of attributes connected to the classes can be extended by including information on dark matter content and dark matter halo parameters. We note that, in conjunction with the WALLABY HI survey, Australian National University's SKyMapper will provide a large homogeneous optical/NIR data set suitable for mass modelling. Future work also includes studying the effect of different parameterizations of the rotation curve, assessed from observational data and simulations in which dark matter parameters are known.

These investigations will provide significant information for selecting the most appropriate demarkations between classes within the continuous distribution.

A robust kinematic classification scheme would be useful for finding characteristics shared by galaxies of similar kinematics and to search for relationships between these attributes. Conversely, knowing only the shape of the rotation curve of a galaxy, a well-calibrated classification scheme could provide information on which properties are likely to accompany galaxies within this rotation curve shape. By extension, this could be used to, for example, collect a sample of galaxies for investigating specific characteristics. Additionally, incorporating halo parameters into our classification scheme could help illuminate the strength of the role played by the gravitational potential in galaxy evolution and star formation efficiency.

## 8. Acknowledgements

We are grateful for the help provided by Jason Fiege in deriving the kinematic parameters for 15 of the galaxies using his HI modelling software GalAPAGOS, as well as helpful comments and discussions. Three-dimensional visualisation was conducted with the S2PLOT programming library [2]. We are grateful to Chris Fluke for his help with creating the s2plot figures. This work has used data from the NVSS [20] and HyperLeda [57] catalogues as well as the NASA/IPAC Extragalactic Database (NED) which is operated by the Jet Propulsion Laboratory, California Institute of Technology, under contract with the National Aeronautics and Space Administration. Many thanks to the anonymous referee for helpful comments.

# Appendix A. Rotation curve fits

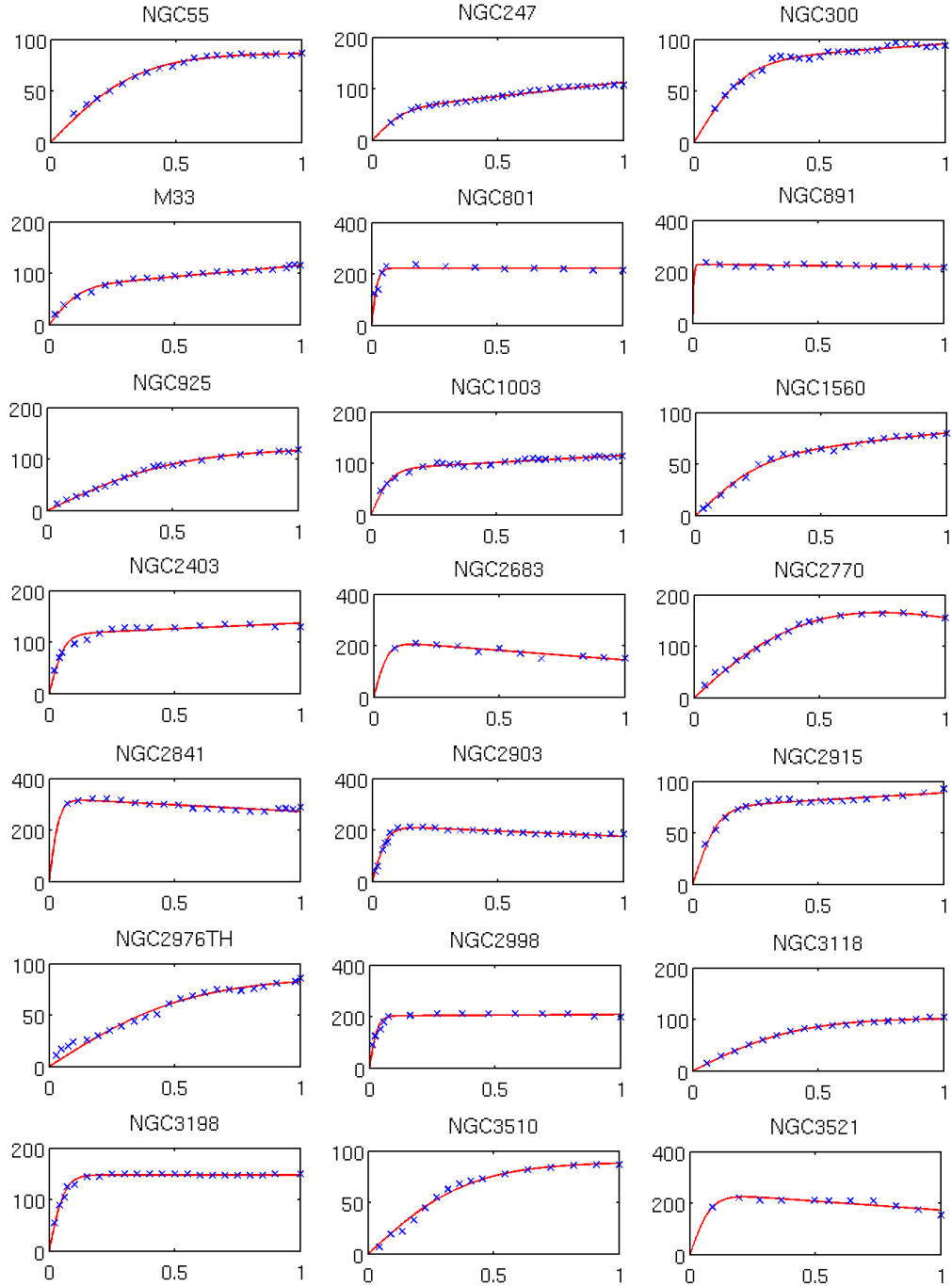

Figure A.21: The rotation curve function in eq. 1 is fitted to the rotation curve data points for the galaxies listed in tables 3, 4 and 5. The horizontal axis is the radius, normalised to the last measured datapoint ($r_{out}$, see column 6 of Tables 3, 4 and 5)). The vertical axis is the velocity in units of km/s.

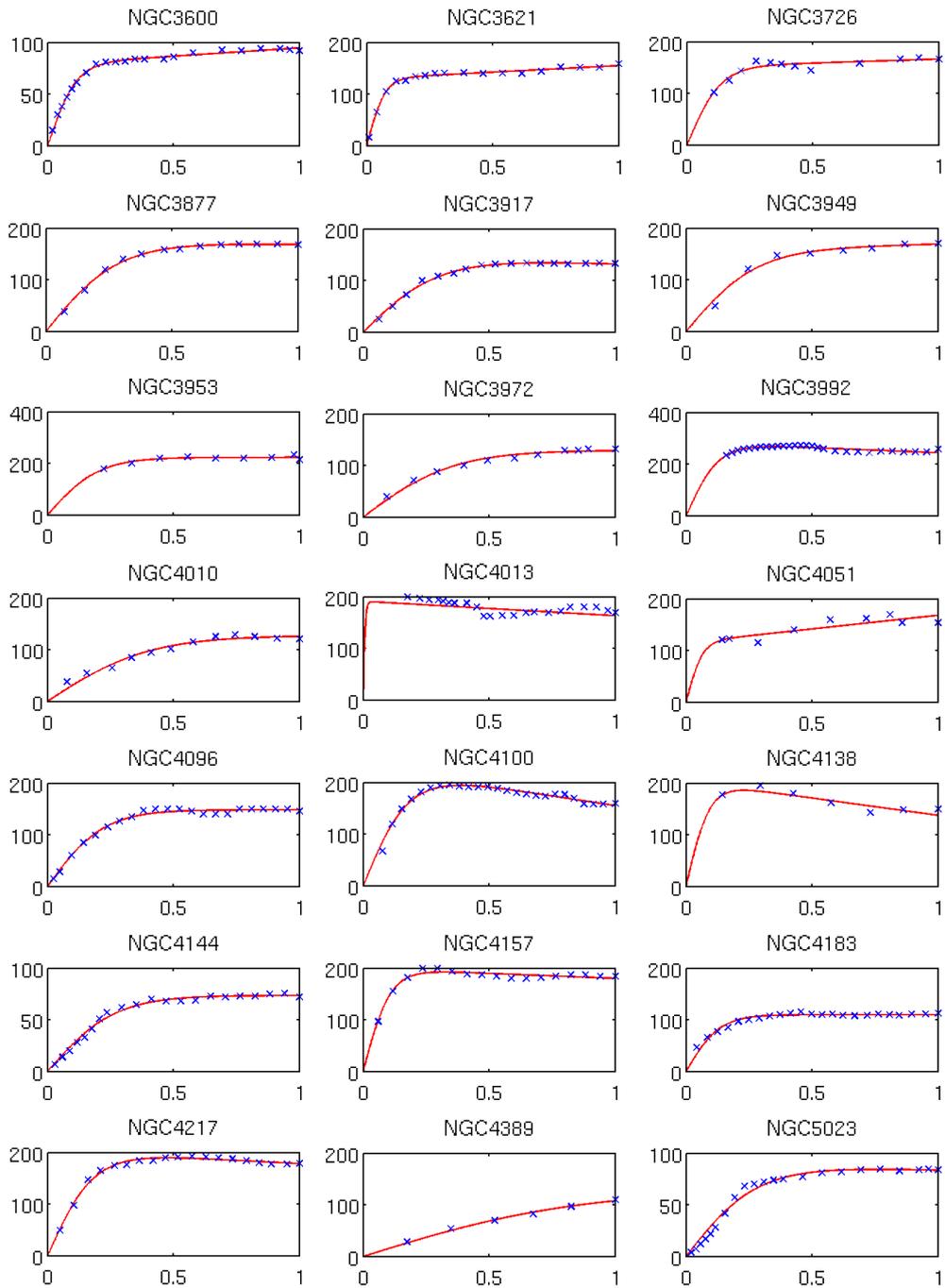

Figure A.21 (continued)

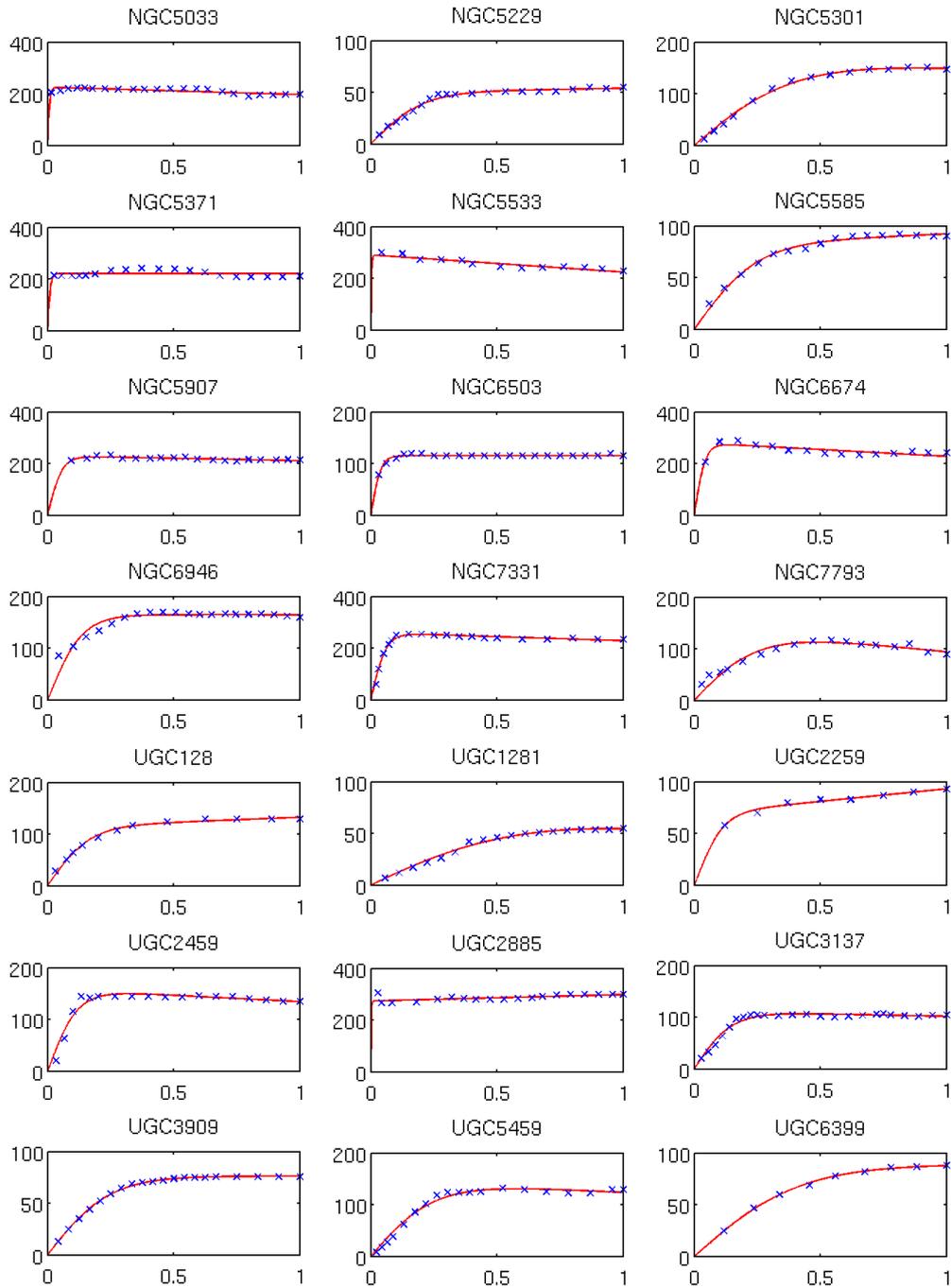

Figure A.21 (continued)

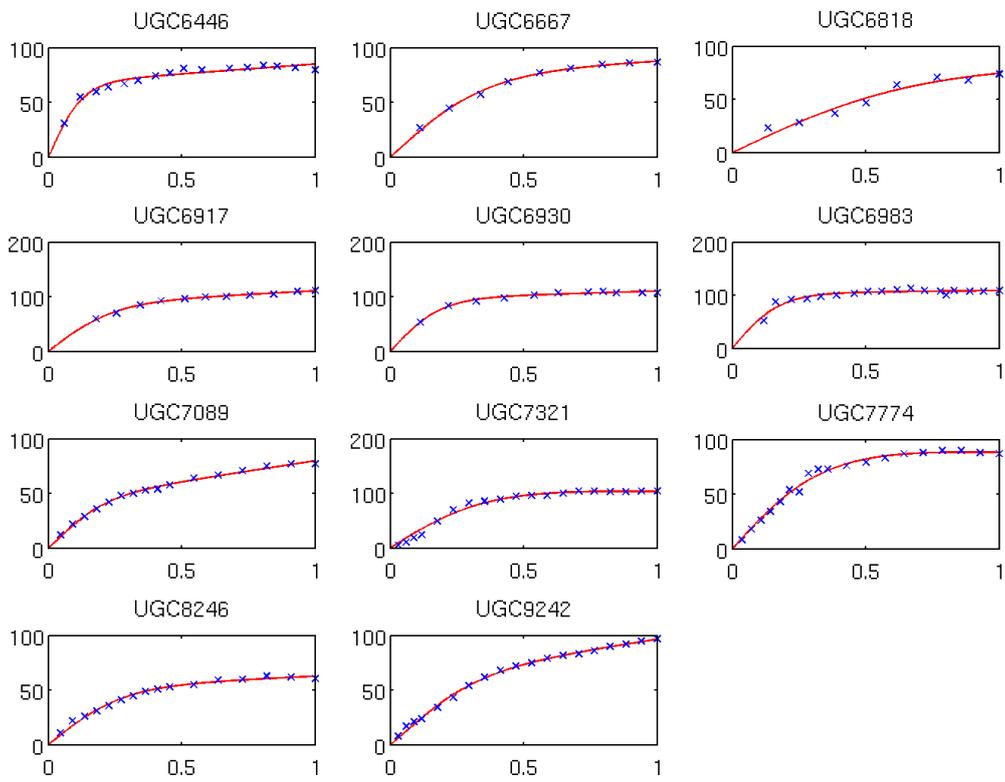

Figure A.21 (continued)